# The Organization of Strong Links in Complex Networks

Sinisa Pajevic[1] and Dietmar Plenz[2]

[1]*Mathematical and Statistical Computing Laboratory, DCB/CIT, NIH*

[2]*Section on Critical Brain Dynamics, LSN, NIMH*

Pages:          26

Text:           4673 words

Figures/Tables: 5

Suppl. Material: 8 Figures, 1 Table

Acknowledgement: We thank Marian Boguna, Hemai Parthasarathy, and members of the Section on Critical Brain Dynamics, NIMH, NIH, for constructive comments during this work. We also thank Shan Yu (NIMH) for providing some of the monkey data and Jeff Alstott (NIMH, Cambridge U) for discussions and Matlab implementation of one of the social network models. This work was supported by the NIH Intramural Research Program of the NIMH and the DCB/CIT.

**Many complex systems reveal a small-world topology[1,2] which allows simultaneously for local and global efficiency in the interaction between system constituents [3-5]. Here, we show that strong interactions in complex systems, quantified by a high link weight, support high network traffic across clustered neighborhoods[1,6]. For brain, gene, social, and language networks, we found a local integrative weight organization in which strong links preferentially occur between nodes with overlapping neighbourhoods with the consequence that globally the clustering is robust to removal of the weakest links. We identify local learning rules that establish integrative networks and improve network traffic in response to past traffic failures. Our findings identify a general organization for complex systems that strikes a balance between efficient local and global communication in their strong interactions, while allowing for robust, exploratory development of weak interactions.**

Networks as diverse as those linking scientific collaborations and those connecting the U.S. electrical power grid are characterized by small-world topology [1,2,7]. In the brain, this topology captures the organization of neural connections at different spatial scales and in various species [8-12], including the structure of spontaneous neural activity characterized by neuronal avalanches [13-15]. In these sparse networks, most nodes are only separated by a few links and nodes are highly clustered, that is, neighboring nodes are very likely to be connected themselves, quantified by a high clustering coefficient, $C$ [7,16,17]. This enables complex systems to simultaneously achieve both global and local efficiency in the interactions of their components [3-5].

In most real world networks, a gradation of interactions exists, commonly quantified by the link weight, $w$ [1,6], which reflects important functional properties such as capacity in transportation routes and communication networks, strength of friendships in social networks, or memories reinforced in brain networks. Recently, many features of weighted networks have been studied, e.g. the relationship between the node degree and node strength [18,19], pair-wise node correlations [20], and dynamical properties [21,22], but some of the earliest findings regarding the relationship between weights and network topology were observed in social networks four decades ago [23,24]. In the seminal work by Granovetter [23], it is stated "that the degree of overlap of two individuals' friendship networks varies directly with the strength of their tie to one another". Thus, strong links are found between nodes with highly overlapping neighbourhoods, a principle that was recently confirmed in mobile phone communications [25]. Here we extend this finding to other complex networks, in particular brain, gene, and human interaction networks. On the other hand, some real networks and network models displayed the opposite behaviour where strong links tend to connect non-overlapping neighbourhoods. This indicates a general principle in weight organization based on local network properties for which the consequences on global network properties are currently not known.

We therefore analyse the relationship between clustering and the weights, both locally and globally, and hence, focus on small-world networks. We treat the neighborhood overlap as a local measure of clustering, and thus define for each link and its two end nodes, the *link clustering coefficient*, $C_L$, as

$$C_L = \frac{n_C}{n_T}, \qquad (1)$$

where $n_C$ is the number of common neighbours and $n_T$ is the total number of neighbouring nodes excluding the end nodes. For directed networks, we use outgoing links for neighborhood definition if not stated otherwise (see also Suppl. Material and Fig. S1). We quantify the relationship between $C_L$ and link weight, $w$, by the correlation coefficient, $R_{C_L}$, and visualize the trend of the *excess link clustering*, $\Delta C_L = C_L^{orig} - C_L^{DSPR}$ vs. weight rank, rank 1 being the smallest weight. Here, $C_L^{DSPR}$ is the link clustering coefficient of the node degree sequence preserving randomized controls (DSPR)[26] and corrects for the overlap contributed by the node degree distribution. In figure 1a, we plot $\Delta C_L$ versus weight rank for three directed networks derived from neuronal avalanche activity in two different types of organotypic neuronal cultures and in the pre-motor cortex in awake macaque monkeys [13-15]. In these networks, weights represent the spontaneous neuronal activity flow between different nodes, i.e. sites in the neural tissue [15]. The steep positive trend we observe, and hence, positive $R_{C_L}$, demonstrates that activity propagates preferentially between nodes with highly overlapping neighbourhoods.

Similar results were obtained for networks of the structural and functional organization of the human cerebral cortex [8] (Fig. 1b), both describing the connectivity between ~1000 cortical regions of interest (nodes) distributed over 67 functional cortical areas. The weights in these two networks are based on axonal fiber density, identified using diffusion spectrum imaging (DSI), and the correlation strength, derived from the ongoing 'resting state' cortical activity using fMRI [8], respectively. In figure 1c, we show similar results for gene regulation networks derived from human and mouse gene expression data [27]. The weights in these networks measure the degree of regulation between two genes. In order to have computationally manageable link analysis, links lower than a threshold of 0.08 were discarded. We point out that similar results were obtained for gene sub-networks containing a smaller and randomly chosen subset of the original nodes (see Suppl. Fig. 2). A comparable weight organization was also found for two co-appearance "social" networks (Fig. 1d), a movie actor collaboration network (N=54K) and the network of characters in the chapters of the novel "Les Miserables" (N=77). In two language networks, consisting of the Reuters News 9/11 network (N=13.3K), and the directed words free-association network (N=10.6K) in which weights represent the co-occurrences of words in news articles and the number of subjects that associated a source word to a target word, respectively, $R_{C_L}$ was positive, but with small $\Delta C_L$ (Fig. 1e).

These results demonstrate that the local weighting rule relating the clustering (or overlap) and the strength of the links, first observed in social networks [23,24], has a more general validity, and is found to hold strongly in the above mentioned biological and social networks. A number of other networks, however, revealed a less positive or even negative trend between neighbourhood overlap and link strength. The anatomically well characterized neural networks of the worm *C. elegans* showed negligible trend, i.e. $R_{C_L} = 0$. Regarding transportation networks, we found positive $R_{C_L}$ for traffic between 500 US airports, and negative $R_{C_L}$ for the US-Air transportation network [1,18], that is strong links preferentially occur between non-overlapping neighborhoods. Similarly, physics author collaboration networks in which weights reflect the number of papers co-authored, normalized by the number of authors on each paper, revealed negative $R_{C_L}$ (for further details on all networks see Suppl. Methods and Table I therein).

For convenience, we define the weight organization with significantly positive $R_{C_L}$ as *integrative* due to the tendency of strong links to connect nodes with overlapping neighbourhoods. Conversely, networks with negative $R_{C_L}$, in which strong links connect non-overlapping neighbourhoods, are defined as *dispersive*. Zero $R_{C_L}$ defines *neutral* weight organization.

**Robustness of clustering to loss of weak links in integrative networks**

The robustness of a network to perturbations, e.g. loss of its nodes, has been an important aspect of network function [28]. Here, we explored the hypothesis that the local integrative weight organization in real world networks is accompanied with the robustness of global clustering properties to the loss of weak links. We used pruning analysis to characterize the change in network topology upon successive removal of the weakest (*bottom*-pruning), or strongest (*top*-pruning) links. For the neuronal avalanche networks described above, we found that, even when a large fraction $f$ of the weakest links was removed, the average clustering coefficient in the network remained high, and the *excess clustering*, defined as $\Delta C = C^{orig} - C^{DSPR}$ remained fairly constant (Fig. 2a; *solid lines*). Here, the excess clustering corrects for the trivial appearance of clustering in finite size networks and converges to $C$ for very large sparse networks. In contrast, removing even a small fraction of the strongest links, i.e. top-pruning, readily destroys clustering in avalanche networks (Fig. 2a; *dashed lines*). Similar results were found for other integrative networks such as the brain, gene, social and language networks (Fig. 2b – f). In contrast, the opposite trend for $\Delta C$ was found in dispersive networks such as the US-Air transportation network (Fig. 2g), which were robust to top- but not bottom pruning. We quantified the difference in constancy of $\Delta C$ between bottom- and top-pruning, by the measure $M$, which ranges between -1 and 1 (see Methods). It is positive for networks in which $\Delta C$ is robust to bottom- but not top-pruning and negative when the opposite is true. $M$ is zero or small for networks that show no robustness or no pruning asymmetry such as collaboration networks (Fig. 2h). The scatter plot of $M$ versus $R_{C_L}$ in figure 2i indicates strong correlation between local integrative weighting and global robustness to loss of strong or weak links for the real networks studied here (R = 0.82). For directed networks, these results on $\Delta C$ and $\Delta C_L$ were independent of whether neighbourhood was defined using outgoing, incoming or all links of a given node (Suppl. Fig. S3).

**Basic models of weight-clustering relationships**

To fully appreciate this particular weight-clustering organization shown in figures 1 and 2, we first compare it to the case when weights are independent from any topological features. Therefore, the correlation $R_{C_L}$ is zero and $\Delta C_L$ shows no trend with respect to link rank. We show analytically in Methods that for independent weights $\Delta C$ decreases linearly for either pruning direction from the initial value $\Delta C_0$ to zero, i.e., $\Delta C(f) = \Delta C_0(1-f)$ and consequently $M=0$. Indeed, in simulations for the directed Ozik-Hunt-Ott growing network (OHO[29]; see Suppl. Mat.) and Watts-Newman (WN) [30] network with randomly assigned weights, $\Delta C_L$ is flat and $\Delta C$ decays linearly to zero for both top- and bottom-pruning (Fig. 3a).

Next, we compare our results with so called Class II networks [19] in which weights are positively correlated with node degrees. An example is the world airline network[18], for which the link weights are related to the end node degrees by,

$$w_{ij} \sim (k_i \cdot k_j)^\theta, \qquad (2)$$

with $\theta = 0.5 \pm 0.1$. We implemented equation 2 in OHO and WN topologies which resulted in networks robust to the loss of its strongest but not weakest links, and in which traffic occurs preferentially between non-overlapping neigborhoods (Fig. 3b). Thus, $\Delta C_L$ decreases with higher $w$, $R_{C_L}$ is negative, and $\Delta C$ remains high for top-, but not bottom-pruning, yielding negative $M$.

We emphasized earlier the local interaction of clustering and weights in integrative networks. To study this further, we created a weight organization model in which we used a different local measure of link clustering that is not directly based on overlap. We assigned link weights to be proportional to the product of the clustering coefficients $C_i$ and $C_j$ of its end nodes,

$$w_{ij} \sim C_i \cdot C_j. \qquad (3)$$

Indeed, its implementation on OHO and WN topologies led to integrative networks with positive $R_{C_L}$ and $\Delta C$ which are robust for bottom-, but not top-pruning (Fig. 3c), as observed in brain, gene, and human networks (Figs.1, 2; see Suppl. Mat Fig. S4 for separation of $C^{orig}$ and $C^{DSPR}$; Fig. S5 using 'all' neighborhood definition). For comparison with real networks, all three models were added to figure 2i.

**Strength of the weak links is their randomness**

Granovetter's work on the "strength of weak ties" and recent work on mobile phone communication [25] demonstrate that weak links serve a cohesive function in complex networks more so than strong links. This could reflect a specific, i.e. targeted organization for weak links that is missed by our definition of integrative, neutral, and dispersive networks. In order to quantify the cohesiveness of the network and its dependence on link weight, we therefore study the reduction in the relative giant component, rGC, during bottom- and top-pruning for our real networks in figure 1. To probe whether the observed cohesiveness arises from targeted weight organization, we compare the change in rGC for both pruning direction to that obtained when pruning links randomly. In figure 4a, we show for the fMRI brain and human gene 1 networks that removal of weak links, i.e. bottom-pruning, reduces the rGC faster compared to random controls in line with the targeted, non-random organization of strong links outlined in the previous sections. In contrast, the change in rGC when pruning from the top, did not differ much from random removal of links (Fig. 4a). We quantify the area between the random control and each pruning direction and show that these findings were true for most integrative networks (Fig. 4b). We conclude that the cohesive character of weak links in the real world networks simply reflects their random nature, rather than targeted placement, further supporting our emphasis on weight organization based on strong links.

**Local dynamical learning rules create integrative and dispersive networks**

Highly clustered neighborhoods with strong links, as found in integrative networks, are known to trap the flow of information [25], hence the 'strength of weak' ties in increasing global efficiency in communication [23]. The question then arises whether integrative networks can alleviate such neighborhood trapping without relying on the random organization for weak links (see Fig. 3), since having strong links in such clustered neighbourhoods can only increase a chance of escape. We demonstrate that this is indeed the case by developing a dynamical model in which local learning rules adaptively change weights in response to past traffic. Using an OHO topology with random weight assignments, traffic was initiated at a randomly selected node and directed probabilistically to future nodes with link weights linearly scaled into probabilities of node activation. This establishes critical branching process dynamics in which one active node leads on average to one active node in the near future. These dynamics serve as a good model for the propagation of avalanche activity in brain networks or other probabilistically propagating traffic that neither grows exponentially nor terminates prematurely. Thus, sequences of activated nodes could span many cascading steps[15], but nodes could only be active once within a cascade and remained refractory until the cascade ended. This behavior is observed experimentally for neuronal avalanches in brain networks and, in general, restricts our exploration to non-cyclical network traffic. After each cascade, the weights of the links between nodes participating in successive time intervals, i.e. cascading steps, were incremented according to different rules (see Methods).

In figure 5a, we show that integrative networks are robustly established when the weight increments are limited to the last step in a cascade. In figures 5c,d we plot the time progression of the parameters $R_{C_L}$ and $M$ during learning and show that this behaviour is observed independent of cascade length (*solid colored lines*). In contrast, limiting learning to any particular pre-defined step beyond the very first link, establishes dispersive networks because longer cascades in networks with randomized weights will reflect the existing degree distribution (Eq. 2; Fig. 5b – d). Learning only at the 1$^{st}$ step, which follows the random initiation of cascades, maintained the initial, neutral weight organization (Fig. 5b – d).

We studied the learning of integrative and dispersive weight organization further by tracking the properties of the cascade termination nodes. As expected, initially, cascades tended to end in neighborhoods of highly clustered nodes quantified by the high correlation between the clustering coefficient of a node and frequency of its participation in cascade termination sites, $R_{C-TN}$ (Fig. 5e). Importantly, the last step learning, instead of exploring alternative routes in the network, directs more future traffic to failure sites by specifically increasing weights for links pointing to terminating nodes in highly clustered neighborhoods. This eventually made those nodes passable for traffic (Fig. 5e, arrowheads), while retaining the integrative weight organization (cp. Fig. 5c,d,e). This improvement in flow was absent in dispersive networks where clustered neighborhoods remained cascade termination points throughout learning (Fig. 5f). This finding was extended to supercritical branching process dynamics, where one node on average activates more than 1 future node (Fig. 2i), whereas cascades in subcritical dynamics failed to reach sufficiently often clustered neighborhoods (Fig. 2i). Similar results where obtained using Watts-Newmann network topology (*data not shown*).

## Comparison with other weighted network models

Finally, we examined several network growth models (GM) with evolving weights. The first two models were originally introduced for networks with preferential attachment [6,31,32]. Because preferential attachment produces scale free networks with no excess clustering, we applied the corresponding weight assignment schemes to OHO growing network topologies to obtain a large $\Delta C$. Neither network model showed integrative properties (Fig. 2i; GM1 OHO, weights based on [32] was dispersive; GM2 OHO, weights based on [6,31] was neutral; see also Suppl. Fig. S6). A third growth model[33], with local weighting and growing rules motivated by social network dynamics and the results of Granovetter, resulted in integrative weight organization, i.e. positive $R_{C_L}$, but low robustness to top-pruning, i.e. small $M$ for a wide range of the model parameters (Fig. 2i; GM3).

## Discussion

Here we identified several important properties of weighted complex networks that are based on the interactions between the clustering and link weights. Earlier findings by Granovetter and colleagues [23,24] have related neighbourhood overlap to link weights between nodes in social networks, which suggests that the level of communication between two people positively correlates with the number of the friends they share [25]. We generalized this local interaction between weights and the clustering and extended its validity to other complex networks, in particular, biological and human interaction networks. We note that this local, link based measure of clustering is essentially equivalent to the edge clustering coefficient [34,35] based on the number of triangles passing through a link, but normalized differently.

We then showed that this local rule is accompanied by globally robust clustering properties in most real world networks, i.e. the robustness of their clustering to a significant loss of their weakest links. Specifically, brain networks of mammalian cortex, gene regulation networks in human and mouse, and some types of social and language networks were quantified by high positive values of $R_{C_L}$ and $M$, respectively. For most real world networks, we found the measures $M$ and $R_{C_L}$ somewhat to be related, since low link clustering for the weakest links also implies less impact to the average clustering coefficient when they are removed. However, the constancy of $\Delta C$ and positive $R_{C_L}$ are different concepts and do not imply high value of each other as demonstrated by their low correlation in some weighting models (see Suppl. Fig. S7). For example, the Kumpula model [33], which essentially implements the Granovetter rule of neighboring overlap, leads to moderately high values of $R_{C_L}$ and $Q$, but $M$ is low for a wide range of model parameters, that is robustness to loss of weak links is weak. We suggest that the high correlation between $R_{C_L}$ and $M$ found for real world integrative networks conveys a particular functional advantage during network growth and development. Specifically, it allows for the rewiring and dynamical exploration of new, weak connections without undermining a network's functionality, which is embedded in the clustering of its strong links.

Our results also indicate a general relationship between local clustering and link weights, as demonstrated by the strong correlation between link clustering and clustering assigned to each link derived from the node clustering coefficients of its end nodes (R=0.73; see also Suppl. Mat. Fig. S8a). We note that the integrative weight organization primarily depends on the weighting model and cannot be explained by purely topological measures. We tabulated many of the topological properties of the observed networks, but none of them correlated significantly with high $R_{C_L}$ or *M*. For example, there was no correlation between the assortativity and $R_{C_L}$ or *M* (R=0.13, R=-0.05) respectively across all real world networks. On the other hand, the global measure of modularity, Q [36], was found to be weakly correlated with our local measure $R_{C_L}$ (R=0.35; Supp. Mat. Fig. S8b). Nevertheless, $R_{C_L}$ and *M* show much greater mutual correlation (R=0.82) than either has with *Q*. Objections that $R_{C_L}$ and *M* are inherently related to each other by their definition hold as well for *Q*, as defined in [36], since the weights themselves define the modules. It was suggested previously [33] that the local integrative weighting leads to higher modularity in networks. Here, we emphasize a strong connection between the integrative weighting and robust clustering. However, it is currently not known, whether the high values of the two global measures *M* and *Q* emerge in real world networks as epiphenomena of the local integrative nature, or whether the local weight adjustments optimize these global networks characteristics.

The division into integrative and dispersive networks requires that a common interpretation of weights is used for all networks, as any inverse transformation of weights would switch the classification. Here we presumed that link weights quantify traffic, flow, intensity, or any other measure of increased communication or interaction between a pair of nodes. Using the weight rank instead of actual weights makes any monotonic transformation of the weights irrelevant and thus reduces the sensitivity of our results to the precise nature of the weighting.

Our learning model also clearly demonstrates that clustered neighbourhoods in integrative weight organization can carry high traffic and do not necessarily stall traffic or trap information flow, which supports efficient communication across the network. Integrative networks are established based on activation history, if adjustments are limited to the last step of the propagation, which tags and removes information trapping. This "learning at the last step" paradigm is similar to temporal difference learning, a widely used rule in artificial intelligence that links sensory input to desired outcome [37]. In neuroscience, it bears great similarity with reward-mediated learning, in which the last step in a sequence of actions taken, i.e. nodes activated, is rewarded given the desired outcome [38]. Importantly, this learning rule does not require specific global information about the network despite dynamically reconfiguring the network as a function of past activity. Accordingly, in networks with high *M*, weak links can be established and modified without compromising the already existing robustness and functionality of the overall network. We suggest that this provides networks with the flexibility to dynamically explore new configurations. For example, during cortex development, weak neuronal connections are constantly formed, removed, or strengthened depending on the activity that occurs between neurons. Integrative weight organization potentially enables neural systems to learn new memories without detrimentally affecting old ones stored in strong connections.

These results suggest that integrative and dispersive weight organization described for real networks captures the targeted organization of strong links that emerge from a random network of weak links.

## Methods

**Link Clustering Analysis.** Since the correlation $R_{C_L}$ between $w$ and $C_L$ is a linear measure, we also studied the trend of link clustering with respect to weight rank. Links ordered by their weight rank were block-averaged to obtain $\langle C_L^{orig} \rangle(i)$ for the $i^{th}$ block, $i = 1,\ldots,10$. We similarly obtain $\langle C_L^{DSPR} \rangle$ from DSPR controls, which show no trend, and subtract this constant offset to obtain the *average excess link clustering* for each block, $\langle \Delta C_L \rangle(i) = \langle C_L^{orig} \rangle(i) - \langle C_L^{DSPR} \rangle$.

**Pruning Analysis.** We studied network topology as a function of the fraction $f$ of the weakest (bottom-pruning) or strongest (top-pruning) links removed. Thus, weights in our pruning analysis mainly serve as labels for link ordering, allowing for easier comparisons between different weighted networks since any monotonic transformation of the weights does not affect our pruning results. The order of removal for links with identical weights was randomized.

Many networks maintained high and approximately constant $\Delta C$ for a particular pruning direction, which we defined as *robust excess clustering* (REC) and quantified using the inverse of the coefficient of variation (CV) of the $\Delta C(f)$ measured across 10 values of $f = [0, 0.1, \ldots, 0.9]$. To reduce large variations in the measure when the standard deviation is extremely small, we used a transformation which confines this measure to the range -1 to 1, i.e.,

$$M_{REC} = 2\,\mathrm{ArcTan}\left(\langle \Delta C(f_i) \rangle / \mathrm{SD}(\Delta C(f_i))\right)/\pi \ . \qquad (4)$$

We calculated $M_{REC}$ for bottom- ($M_{REC}^{Bottom}$) and top- ($M_{REC}^{Top}$) pruning profiles, $\Delta C(f)$, and use their difference $M$ to quantify the asymmetry

$$M = M_{REC}^{Bottom} - M_{REC}^{Top}. \qquad (5)$$

$M$ is positive for integrative networks and negative for dispersive networks, whereas it is close to zero for neutral networks.

To quantify the difference in the change of rGC, networks and controls were pruned until all links were removed. The area between the random removal curve and top and bottom pruning curves respectively was integrated. Positive/negative values indicate cohesiveness less/better than random respectively.

**Local Learning Rules.** We studied the weight organization resulting from dynamical learning that occurs during a branching process dynamics [15]. We simulated critical branching process on OHO and WN topologies initiated with uniform or random, but narrowly distributed weights (neutral). The weights, $w_{ij}$, were appropriately scaled to be interpreted as the critical branching process probabilities of the source node $i$ activating the target node $j$[15]. Before the next initiation, the scaling factor, which converts link weights into branching process probabilities was adjusted such that the network dynamics remains critical. After each cascade, we changed weights for the links connecting nodes in successive time intervals (generations) according to

$$w_{t+1} = w_t(1 + p_p), \; p_p = (w_{max} - w_t)/w_{max}, \qquad (6)$$

where $p_p$ (ranging from 0.01 to 1%) is a small percent increase factor and $w_{max}$ is the maximum weight allowed (5 to 500). Importantly, we restrict learning to particular successions, i.e. steps, in the following four ways: a) learning only at a particular step (e.g. $1^{st}$, $2^{nd}$, …); b) learning at the last step of every cascade; c) at a particular step, but only if it also is the last step and d) at all steps without restrictions. Only with learning restricted to the last steps (restrictions b and c) integrative behaviour occurred and over a wide range of parameters until all weights eventually saturate to the maximal value $w_{max}$. Results shown were taken before significant weight saturation occurred.

We quantified termination of cascades by the frequency of the appearance of a particular node in the last time interval of a cascade, i.e. was a terminal node (TN). We calculated the fraction, $f_{TN}$, of all cascades in which a node was a TN and calculated the correlation $R_{C-TN}$ between $f_{TN}$ and the clustering coefficient $C$ across all nodes.

**Analytical results for independent weights.** The pruning of a network in which link weights are independent from topology is equivalent to removing links randomly. Upon removal of the $t^{th}$ link, only the clustering coefficients of the $n_c$ common neighbours of its end nodes (Fig. 1a) are reduced, hence, the average clustering coefficient, $C_t$, changes according to

$$C_{t+1} = \frac{n_c}{N}\left(C_t - \frac{1}{z_t(z_t-1)}\right) + \frac{(N-n_c)}{N}C_t, \tag{7}$$

where $z_t$ is the average degree. In the continuous limit Eq. 7 becomes

$$\frac{dC}{dt} = -\frac{n_c(t)}{Nz(t)(z(t)-1)} \ . \tag{8}$$

When the weights are independent from topology, $n_c(t) = (z(t)-1)C(t)$, and

$$\frac{dC}{dt} = -\frac{C}{Nz(t)} \ . \tag{9}$$

One can similarly obtain a differential equation for $z(t)$, whose solution is

$$z(t) = z_0 - \frac{t}{N} \ . \tag{10}$$

Using Eq. 10 in Eq. 8, we obtain

$$\frac{dC}{dt} = -\frac{C}{Nz_0 - t} = -\frac{C}{t_0 - t}, \tag{11}$$

where $t_0$ is the total number of links in the original network. Solution of this equation is

$$C = C_0(1-f) \ , \tag{12}$$

where $f = t/t_0$ is the fraction of the removed links. The excess clustering is a difference of two clustering coefficients, both decaying with the same rate $1-f$, hence

$$\Delta C = \Delta C_0(1-f). \tag{13}$$

# Figures

**Figure 1** Link clustering in real world networks reveals preferential placement of strong links with respect to the neighbourhood overlap of the corresponding end nodes. **a**, Functional connectivity derived from ongoing neuronal avalanche activity in left premotor cortex of awake macaque monkey (*black*, n=3) and organotypic cortex cultures (*red*, n = 7; *green*, n = 7, externally driven), and. Average link clustering $\Delta C_L$ plotted vs. the weight rank. Note the strong positive trend for link weights to increase with increase in relative neighbourhood overlap of end nodes, i.e. $\Delta C_L$. First rank is smallest weight. **b**, Functional architecture of the human cerebral cortex obtained using fMRI (*black*; n = 5 subjects) and corresponding structural cortex core obtained with DSI (*red*). Strong connections preferentially occur between sites with high $\Delta C_L$. **c,** Gene expression networks derived from human (*black*, *red*) and mouse (*green*) gene expression data. **d,** Social co-appearance networks represented by a movie actors network and the network of characters in the novel "Les Misérables". **e**, A weak positive trend for $\Delta C_L$ characterizes the word association and Reuters 9/11 News network. **f**, Summary analysis for all three networks of *C. elegans* reveals no trend in link clustering. **g**, US air flight network shows a weak, negative trend for $\Delta C_L$ indicating that strong routes preferentially connect airports that serve different destinations, i.e. reduced neighbourhood overlap. Airport passenger network reveals low to slightly positive $\Delta C_L$. **h**, Author collaboration networks with co-authorship weighted by the total number of authors on a paper. Negative trend in $\Delta C_L$.

**Figure 2** The robustness of clustering to loss of their weakest or strongest links in small-world networks and its correlation with link clustering. **a**, Neuronal avalanche networks from awake macaque monkeys (*black*) and organotypic cortex cultures (*red*, *green*). $\Delta C$ remains constant for bottom-pruning (*solid lines*), i.e fraction, $f$, of weakest links pruned, , but not top-pruning (*broken lines*.), i.e. $f$ of strongest links pruned. **b – f**, Robust $\Delta C$ to bottom but not top-pruning also characterizes the human brain, gene interaction, social, language and *C. elegans* networks. **g**, Transportation networks such as the US air flights and airport passenger networks are robust to top pruning, but not bottom-pruning, that is clustering largely depends on weak links. Note that high capacity routes for US air flights are formed between airports with a clustering coefficient below chance. **f**, Summary plot of $M$ vs. $R_{C_L}$ for all networks analyzed in the present study. Brain, gene, social, and language networks are integrative with brain and gene networks exhibiting among the highest positive values of $M$ and $R_{C_L}$. We note that only models (OHO II, GM1) achieve high dispersive characteristics, whereas natural networks like Airline and collaboration networks range from weakly dispersive to neutral.

**Figure 3** Link clustering and pruning analysis for neutral, dispersive, and integrative weight organizations. Simulations are shown for OHO (*black*) and WN (*red*) topologies (n=10 networks, each N=100; <k> = 12, 10 for OHO, WN respectively; see Suppl. Table 1). **a**, Example of neutral weight organization with randomly assigned link weights $w$ (independent of the topology). *Left*: $\Delta C_L$ shows no trend vs. weight rank  *Right*: $\Delta C$ decreases linearly with $f$ for bottom- (*solid*) and top-pruning (*broken*). **b**, In networks with dispersive weight organization, here implemented according to Eq. 2, $\Delta C_L$ is highest for weak links and $\Delta C$ is robust only for top-pruning. **c**, In networks with integrative weight organization, here implemented using Eq. 3, $\Delta C_L$ is highest for strong links and $\Delta C$ is only robust for bottom-pruning.

**Figure 4** The cohesive nature of weak links is grounded in their random organization. **a**, the relative giant component, *rGC*, in fMRI brain and human gene 1 networks drops faster for top pruning (*black*) compared to randomized weight controls (*red*). In contrast, the rGC changes similarly to randomized weight controls for top-pruning. **b**, Summary plots of the difference in *rGC* for bottom- and top-pruning compared to randomized weight controls for all real world networks. The small difference for integrative networks when pruned from the top suggest random organization of weak links. Conversely, the large difference for bottom pruning indicates targeted, non-random organization of strong links.

**Figure 5** Adaptive implementation of integrative and dispersive weight organizations. **a**, An initial random, i.e. neutral, weight assignment (*red*) changes into integrative (*black*) during last step learning (*after*: $10^6$ cascades; OHO topology; N =60; n = 5 realizations; see also Suppl. Table 1). **b**, Learning only at the 1st step (2nd step) results in neutral (dispersive) weight organizations. **c**, **d**, Temporal progression of *M* and $R_{C_L}$ during last-step learning (*solid black*), at any particular step conditioned on it being also the last step (*colored solid*), all-step learning (*dashed black*), and at any particular step (*colored dashed*). Learning at the 1st, 2nd, .., or 5th cascade step (solid lines, L1 – L5), if this step also was the last in the cascade, results in integrative networks. Learning at every 1$^{st}$ step (A1; *red dashed*) maintains neutral networks, while dispersive networks emerge for later steps (A2 – A5) *All*: all-step learning; *Last*: learning at all last steps. **e,** Last step learning enables cascades to break through traffic traps that exist in clustered neighborhoods during early stages of learning. Temporal progression of $R_{C-TN}$ during last step learning (*Top*, integrative). **f**, In dispersive networks clustered neighbourhoods continue to stall traffic throughout learning. Legend in c applies to d – f.

Reference List

1. Amaral,L.A., Scala,A., Barthelemy,M. & Stanley,H.E. Classes of small-world networks. *Proc. Natl. Acad. Sci. U. S. A* **97**, 11149-11152 (2000).

2. Dorogovtsev,S.N. & Mendes,J.F.F. Evolution of networks: From biological nets to the internet and WWW. University Press, Oxford, USA (2003).

3. Latora,V. & Marchiori,M. Efficient behavior of small-world networks. *Phys Rev. Lett.* **87**, 198701 (2001).

4. Petermann,T. & De Los,R.P. Physical realizability of small-world networks. *Phys Rev. E Stat. Nonlin. Soft. Matter Phys* **73**, 026114 (2006).

5. Achard,S. & Bullmore,E. Efficiency and cost of economical brain functional networks. *PLoS Comput. Biol.* **3**, e17 (2007).

6. Barrat,A., Barthelemy,M. & Vespignani,A. Weighted evolving networks: coupling topology and weight dynamics. *Phys Rev. Lett* **92**, 228701 (2004).

7. Watts,D.J. & Strogatz,S.H. Collective dynamics of 'small-world' networks. *Nature* **393**, 440-442 (1998).

8. Hagmann,P. *et al.* Mapping the structural core of human cerebral cortex. *PLoS. Biol.* **6**, e159 (2008).

9. Sporns,O. & Honey,C.J. Small worlds inside big brains. *Proc. Natl. Acad. Sci. U. S. A* **103**, 19219-19220 (2006).

10. Sporns,O. Small-world connectivity, motif composition, and complexity of fractal neuronal connections. *Biosystems* **85**, 55-64 (2006).

11. Eguiluz,V.M., Chialvo,D.R., Cecchi,G.A., Baliki,M. & Apkarian,A.V. Scale-free brain functional networks. *Phys Rev. Lett* **94**, 018102 (2005).

12. Bassett,D.S., Meyer-Lindenberg,A., Achard,S., Duke,T. & Bullmore,E. Adaptive reconfiguration of fractal small-world human brain functional networks. *Proc. Natl. Acad. Sci. U. S. A* **103**, 19518-19523 (2006).

13. Beggs,J.M. & Plenz,D. Neuronal avalanches in neocortical circuits. *J. Neurosci.* **23**, 11167-11177 (2003).

14. Petermann,T. *et al.* Spontaneous cortical activity in awake monkeys composed of neuronal avalanches. *Proc. Natl. Acad. Sci. U. S. A* **106**, 15921-15926 (2009).

15. Pajevic,S. & Plenz,D. Efficient network reconstruction from dynamical cascades identifies small-world topology from neuronal avalanches. *PLoS Comp. Biol.* **5**, e1000271 (2008).

16. Strogatz,S.H. Exploring complex networks. *Nature* **410**, 268-276 (2001).

17. YU,S., Huang,D., Singer,W. & Nikolic,D. A Small World of Neuronal Synchrony. *Cereb. Cortex* (2008).

18. Barrat,A., Barthelemy,M., Pastor-Satorras,R. & Vespignani,A. The architecture of complex weighted networks. *Proc. Natl. Acad. Sci. U. S. A* **101**, 3747-3752 (2004).

19. Bianconi,G. Emergence of weight-topology correlations in complex scale-free networks. *Europhysics Letters* **71**, 1029-1035 (2005).

20. Serrano,M.A., Boguna,M. & Pastor-Satorras,R. Correlations in weighted networks. *Phys Rev. E Stat. Nonlin. Soft. Matter Phys* **74**, 055101 (2006).

21. Restrepo,J.G., Ott,E. & Hunt,B.R. Weighted percolation on directed networks. *Phys Rev. Lett.* **100**, 058701 (2008).

22. Restrepo,J.G., Ott,E. & Hunt,B.R. Characterizing the dynamical importance of network nodes and links. *Phys Rev. Lett.* **97**, 094102 (2006).

23. Granovetter,M.S. The strength of weak ties. *American Journal of Sociology* **78**, 1360 (1973).

24. Davis,J. Clustering and structural balance in graphs. *Human Relations* **20**, 181-187 (1967).

25. Onnela,J.P. *et al.* Structure and tie strengths in mobile communication networks. *Proc. Natl. Acad. Sci. U. S. A* **104**, 7332-7336 (2007).

26. Maslov,S. & Sneppen,K. Specificity and stability in topology of protein networks. *Science* **296**, 910-913 (2002).

27. Gregoretti,F., Belcastro,V., di Bernardo,D. & Oliva,G. A parallel implementation of the network identification by multiple regression (NIR) algorithm to reverse-engineer regulatory gene networks. *PLoS ONE.* **5**, e10179 (2010).

28. Albert,R., Jeong,H. & Barabasi,A.L. Error and attack tolerance of complex networks. *Nature* **406**, 378-382 (2000).

29. Ozik,J., Hunt,B.R. & Ott,E. Growing networks with geographical attachment preference: emergence of small worlds. *Phys Rev. E. Stat. Nonlin. Soft. Matter Phys* **69**, 026108 (2004).

30. Newman,M.E. & Watts,D.J. Scaling and percolation in the small-world network model. *Phys Rev. E. Stat. Phys Plasmas. Fluids Relat Interdiscip. Topics.* **60**, 7332-7342 (1999).

31. Barrat,A., Barthelemy,M. & Vespignani,A. Modeling the evolution of weighted networks. *Phys Rev. E. Stat. Nonlin. Soft. Matter Phys* **70**, 066149 (2004).

32. Yook,S.H., Jeong,H., Barabasi,A.L. & Tu,Y. Weighted evolving networks. *Phys. Rev. Lett.* **86**, 5835-5838 (2001).

33. Kumpula,J.M., Onnela,J.P., Saramaki,J., Kaski,K. & Kertesz,J. Emergence of communities in weighted networks. *Phys Rev. Lett.* **99**, 228701 (2007).

34. Radicchi,F., Castellano,C., Cecconi,F., Loreto,V. & Parisi,D. Defining and identifying communities in networks. *Proc. Natl. Acad. Sci. U. S. A* **101**, 2658-2663 (2004).

35. Serrano,M.A. & Boguna,M. Clustering in complex networks. I. General formalism. *Phys Rev. E Stat. Nonlin. Soft. Matter Phys* **74**, 056114 (2006).

36. Girvan,M. & Newman,M.E. Community structure in social and biological networks. *Proc. Natl. Acad. Sci. U. S. A* **99**, 7821-7826 (2002).

37. Sutton,R.S. & Barto,A.G. Reinforcement learning: an introduction. *IEEE Trans. Neural Netw.* **9**, 1054 (1998).

38. Schultz,W., Dayan,P. & Montague,P.R. A neural substrate of prediction and reward. *Science* **275**, 1593-1599 (1997).

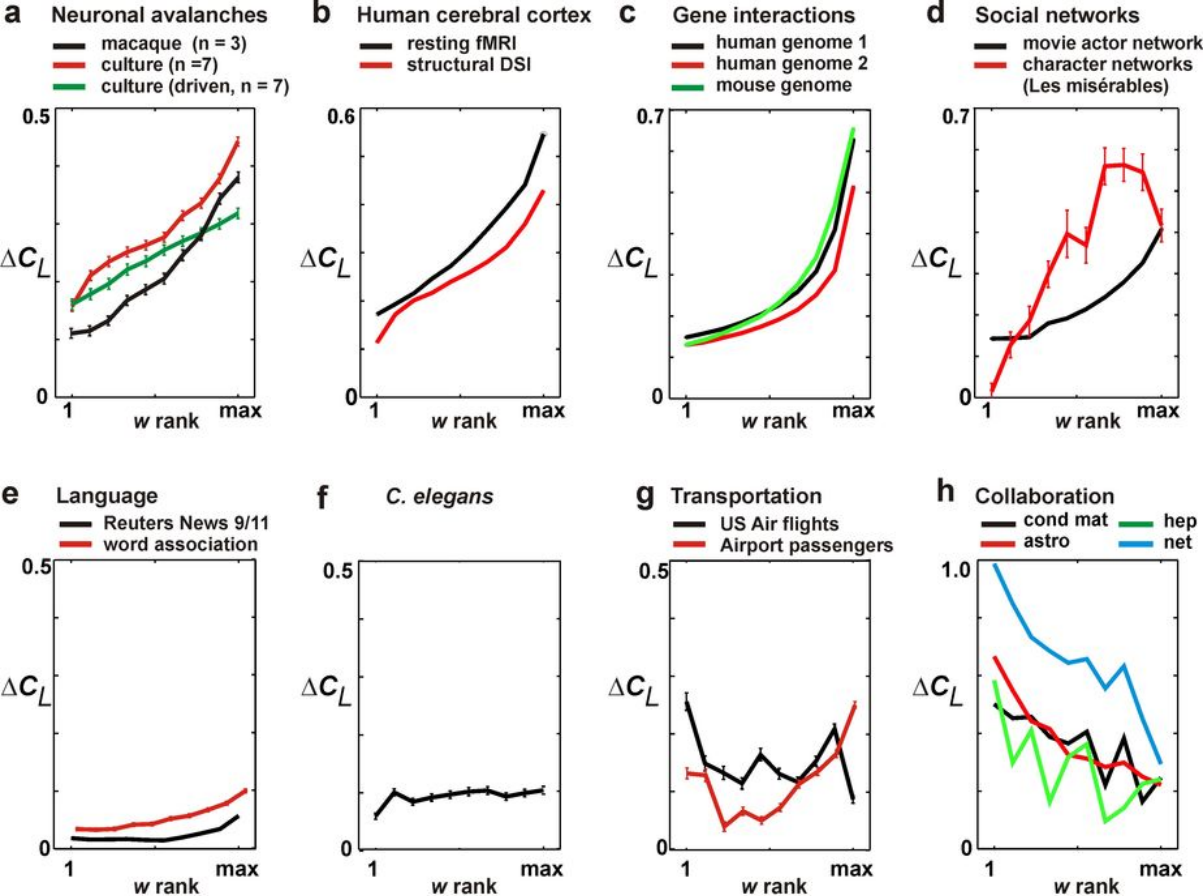

Pajevic & Plenz
Figure 1
2 cols

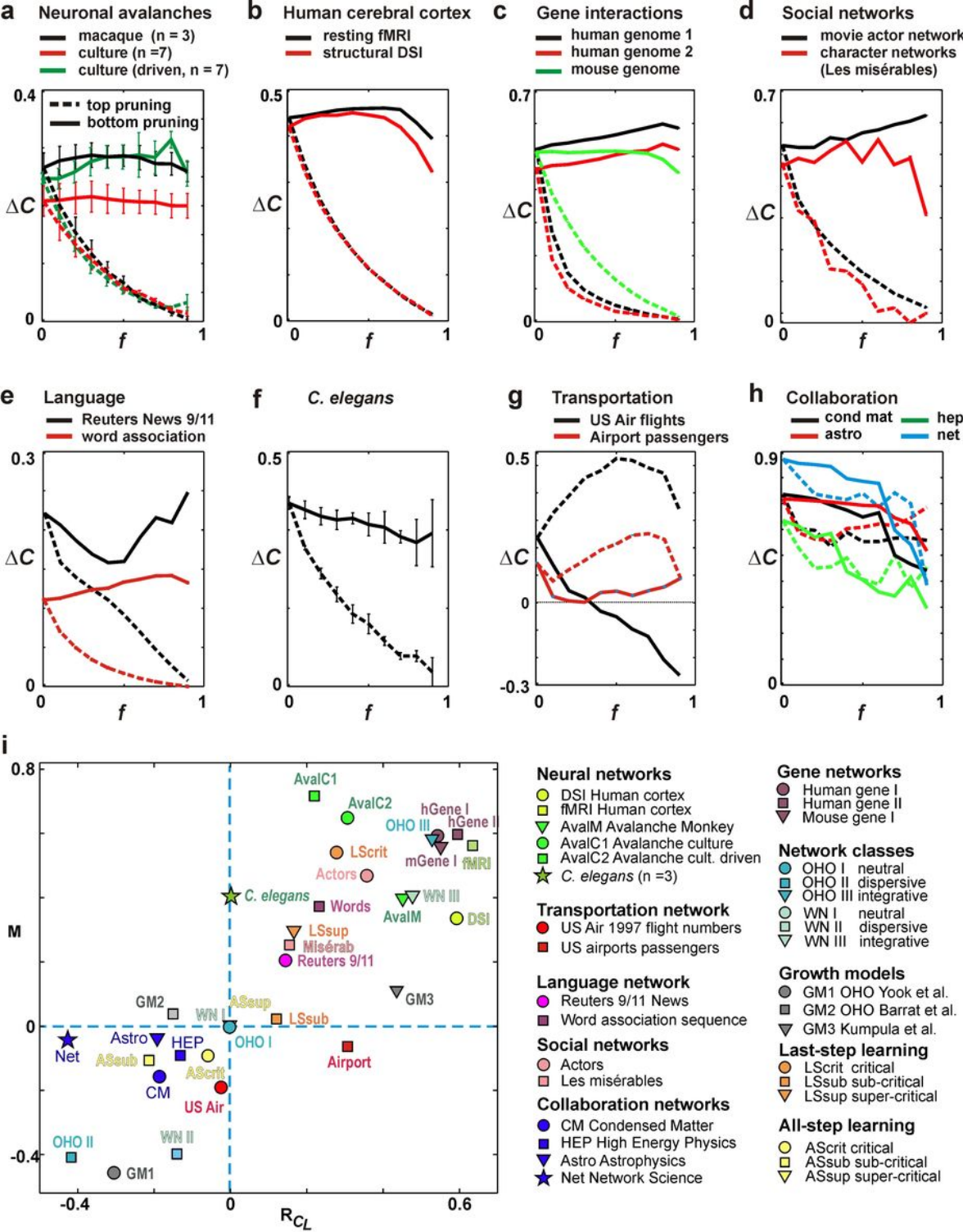

Pajevic & Plenz
Figure 2
2 cols

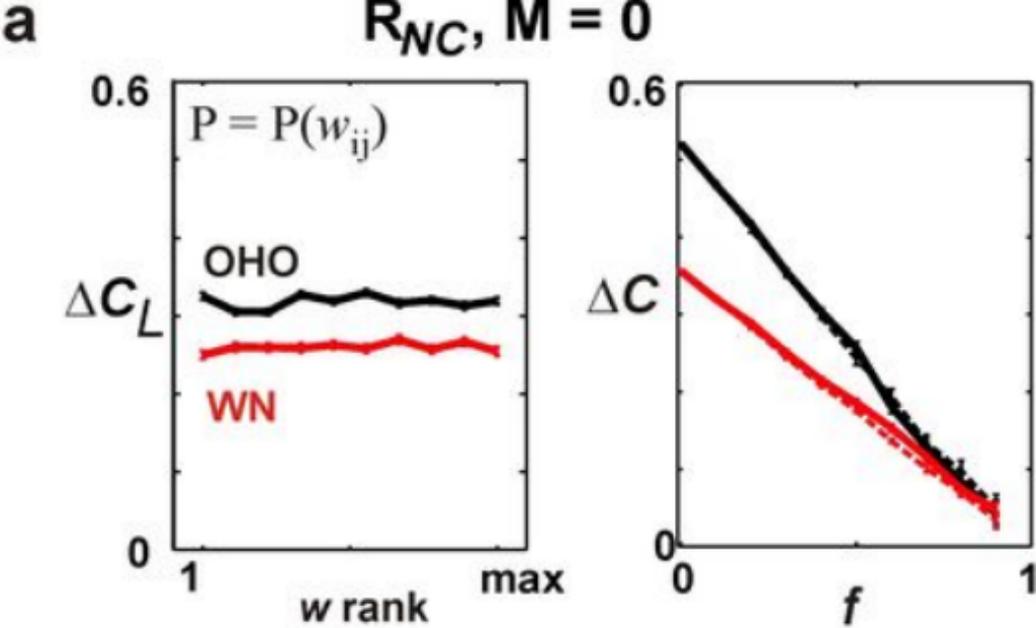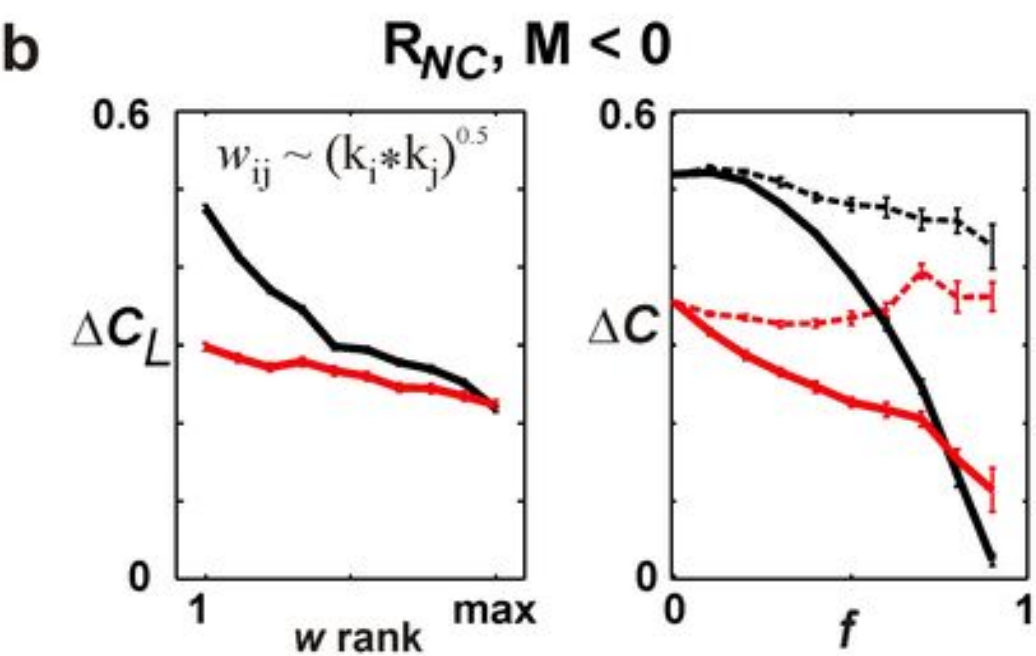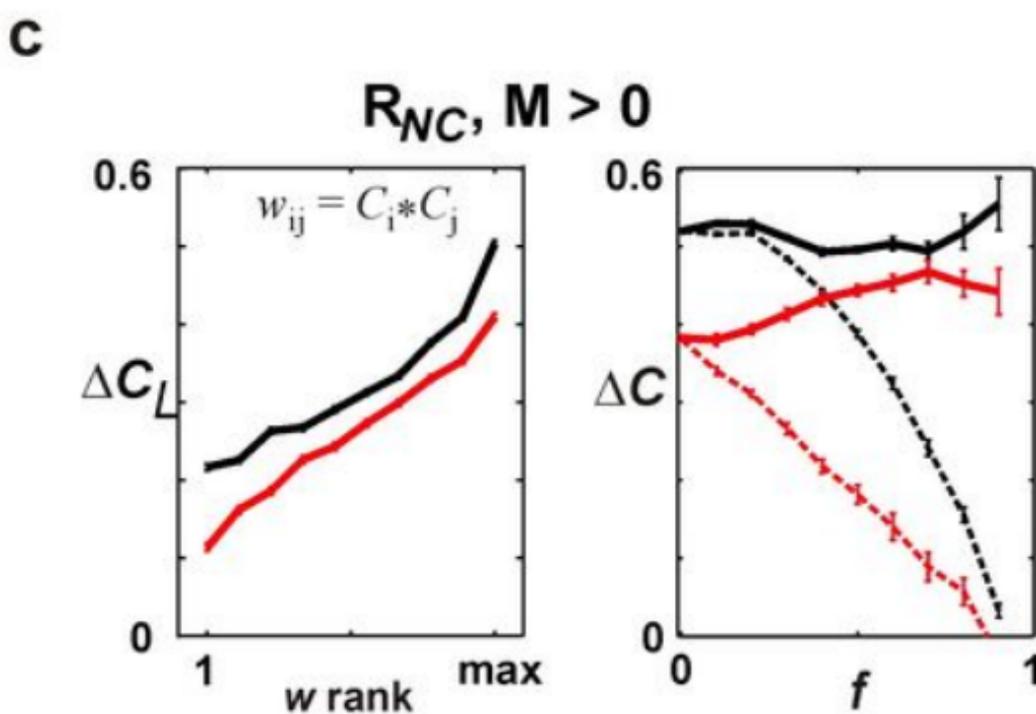

Pajevic & Plenz
Figure 3
1 col

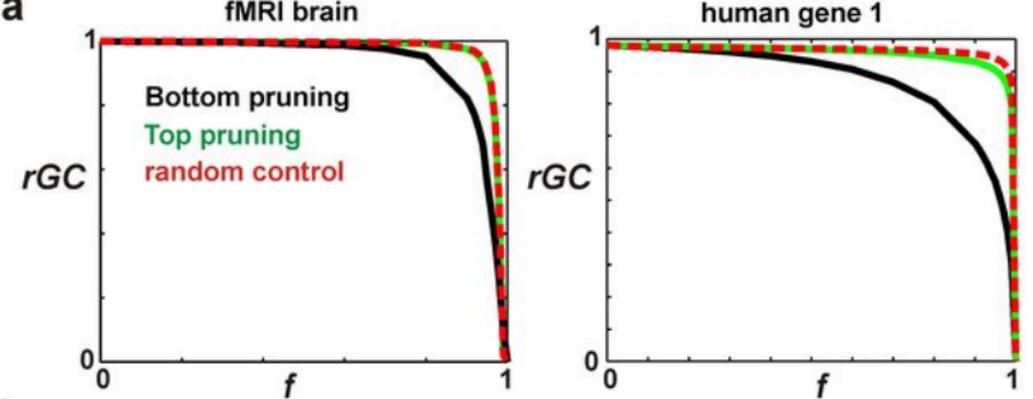
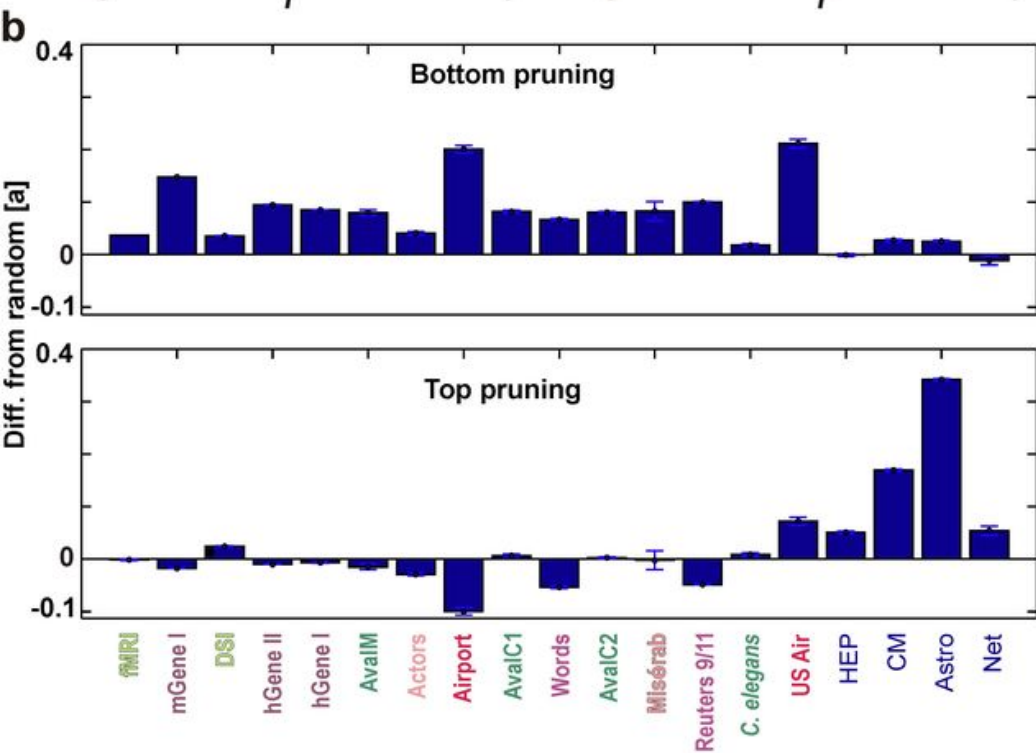

Pajevic &Plenz
Figure4
1.5 col

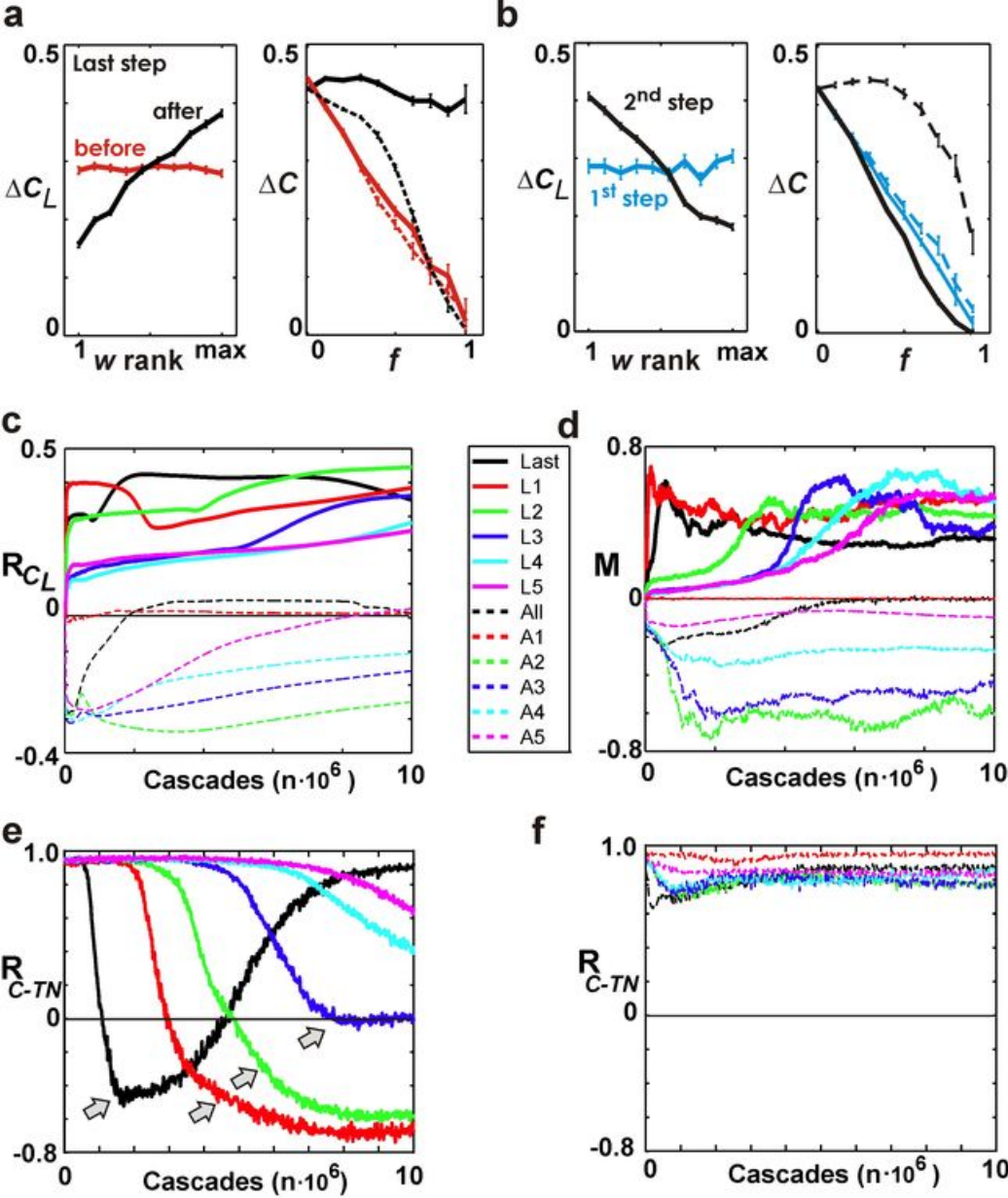

Pajevic & Plenz
Figure 5
1.5 cols

# SUPPLEMENTARY MATERIAL

# The Organization of Strong Links in Complex Networks

Sinisa Pajevic[1] and Dietmar Plenz[2]

[1]*Mathematical and Statistical Computing Laboratory, DCB/CIT, NIH*

[2]*Section on Critical Brain Dynamics, LSN, NIMH*

## 1. Topological measures of clustering in networks

*Node clustering coefficient.*

For each node *i*, the node clustering coefficient, $C_i$, was defined as the probability that an edge between any two of its neighbours exists [1]. This concept is easily extended for directed networks, i.e.,

$$C_i = \frac{\sum_{j,k \in \Omega_i} a_{jk}}{k_i(k_i - 1)}, \qquad (S1)$$

where $\Omega_i$ contains all neighbours of node *i* and $a_{jk}$ is an element of the adjacency matrix so that $a_{jk} = 1$ if a directed link from *j* to *k* exists, otherwise it is zero. The clustering coefficient *C* of a network is the average of $C_i$ for all nodes with degree 2 or greater. For directed networks, $\Omega_i$ and, hence, $C_i$ can be defined based on the '*out*', '*in*', or '*all*' neighbourhood utilizing either the out-going, in-coming, or all of the links of a node respectively. If not stated otherwise, our analysis for directed networks is based on '*out*'-neighborhood.

*Link Clustering Coefficient*

For any directed or undirected link, we define the neighbourhoods for each of its end nodes, either by choosing nodes that it points to (*out* neighbourhood), nodes that point to it (*in* neighbourhood) or without considering direction (*all* neighbourhood), noting that a different type of the neighbourhood can be chosen for the source and the target node. Then, the whole set of pure neighbours (that exclude the source and the target nodes themselves) can be divided into three groups: the common nodes, $n_C$, and two sets that are unique to the source and to the target ($n_U$, see Fig. S1a). In clustered networks the number of common nodes will be much larger than in the equivalent randomized network, and hence we use it as a local measure of clustering defined for each link. More precisely, the link clustering coefficient, $C_L$, for a given link is defined as

$$C_L = \frac{n_C}{n_T}, \tag{S2}$$

Where $n_C$ is the number of common neighbours of the link's end nodes, and $n_T$ is the total number of end node neighbours excluding the end nodes themselves (see Figure S1a). We developed two quantitative measures to study the relationship of link weight $w$ and $C_L$ in weighted networks.

A similar measure of clustering local to the edges, called *edge clustering*, has been defined in ref [2] as the fraction of the triangles passing through an edge out of the total possible number of triangles that could potentially exist between the neighbours. This can be written as

$$C_{i,j}^{(3)} = \frac{z_{i,j}^3}{\min[(k_i-1),(k_j-1)]},$$

where $C_{i,j}^{(3)}$ is the edge coefficient, $z_{i,j}^3$ is the number of triangles passing through the edge (i,j), and $k_i$ and $k_j$ are the degrees of the end nodes of the edge. For undirected networks this measure differs from $C_L$ only in the way it is normalized. However, our measure is a more conservative measure of clustering and our definition enables easier adoption to directed networks by merely changing the definition of neighbourhood. Thus, by taking into consideration the direction of the link between the two end nodes as well as their neighbours, we have examined 5 of the total 9 pairs of neighbourhood schemes (out-out, in-in, all-all, in-out, out-in; see Suppl. Fig. S3). We have also explored many different normalization schemes, including the one used for edge clustering coefficient, none of which changes the nature of correlations with weights significantly.

*Excess Clustering*

In our analysis, we use *excess clustering*, $\Delta C$, which we define as the difference between the clustering coefficient of the original network, $C^{orig}$, and that of an equivalent randomized network, $C^{DSPR}$, with the degree sequence preserved (DSPR [3]),

$$\Delta C = C^{orig} - C^{DSPR}, \tag{S3}$$

The reason for preferring $\Delta C$ over $C$ is that $C$ can indicate high clustering even when it arises trivially from the prescribed node degree sequence or degree distribution. Notably, any complete graph has $C = 1$. On the other hand, for a network to have significant $\Delta C$, some form of targeted connectivity ought to be present in network formation. It is widely accepted that true clustering requires the presence of diverse node groups and some preference of attaching to one group over the others are required [4-7], which in general can arise through the presence of hidden metric spaces[8,9].

Technically, we obtained DSPR networks by repeated random selections of a pair of directed links with distinct source and target nodes, and then switching the target nodes. The number of switches was twice the total number of links. At each pruning level, we estimated $\Delta C$ by averaging over a certain number of randomized networks (with slightly different values of $C^{DSPR}$), which ranged from 2 randomizations for networks larger than 10,000 nodes and up to 20 for the smallest networks..

*WN and OHO network topologies*

In our models, we tested two common small-world topologies. The Watts-Newman topology (WN)[10] is a simpler version of the Watts-Strogatz topology[1]. Long range random links are added to a regular lattice connecting K nearest neighbors without rewiring the existing lattice links. In essence, it is a simple superposition of an Erdos-Renyi network with a regular lattice network. In our implementation, we use K=4 and p=2/N, yielding average degree of 10. The Ozik-Hunt-Ott (OHO)[11] topology is a growing network model which starts with a simple lattice to which new nodes are inserted in between two randomly chosen neighbors and forming links to K nearest neighbors. This model yields a highly clustered network with C ~ 0.7 which is independent of N and exponentially distributed degree distribution. We used K=6, for which the average degree is 12.

## 2. Supplementary Analysis

*Alternative measure of correlation between weights and clustering*

We showed that an integrative weight organization can be introduced in typical small-world topologies by assigning link weights $w_{ij}$ proportional to the clustering coefficients of the corresponding end nodes $i$ and $j$ as in Eq. 3 in the main text. This assignment results in high positive values for $M$ and $R_{C_L}$. In addition, we studied the correlation $R_w$ between the weights in the original network, $w$, and those assigned by the Eq. 3 in the text, i.e.

$$R_w = \text{Corr}(w_{ij}, C_i \cdot C_j) . \quad (S4)$$

We calculated $R_w$ for natural and simulated networks and found, as shown in Suppl. Fig. S8a that $R_w$ correlates with $R_{C_L}$ (R=0.73).

*Correlation between the two measures of weight organization, $M$ and $R_{C_L}$*

We explored empirically the potential correlation between the two measures of weight organization $M$ and $R_{C_L}$. Given that the weighting model in Eq.3 (main text) produces high $M$, but not as high $R_{C_L}$ as one would obtain if $C_L$ was used directly to develop a weighting scheme, we aimed to de-correlate the two measures. More specifically, we applied weights according to

$$w_{ij} \sim C_i \cdot C_j - a C_L^{(i,j)} , \quad (S5)$$

where *a* is a control parameter which we studied for a wide range of values ranging from much smaller than 1 (0.01) to much larger than 1 (100) spaced logarithmically. Our goal was to identify a range of *a* for which the measures had different signs, i.e. fall outside the Upper-Right and Lower-Left quadrants of Fig. 2i. Indeed, as shown in Suppl. Fig. S7, in which the weighting scheme was applied for OHO and WN topologies the measure can take on opposite signs. We also note that for a wide range of negative values of $R_{C_L}$, the two measures were either uncorrelated (OHO) or even slightly anti-correlated.

*Giant Component and Mean Path Length Pruning Analysis*

Results of pruning will also depend on the initial level of sparsity in the network and the network pre-processing. For example, for a fully connected, i.e. complete network, the excess clustering is zero and hence an increase in $\Delta C$ is to be expected. For the fMRI functional network, we established an independent and commonly used threshold, thus declaring the default level of significance in measured correlations. In gene and actor networks, due to their extremely large size we were forced to omit the weakest links. Since the pruned networks exhibited robustness to bottom pruning, we could proceed consistently. We note that care has to be taken that such pre-processing does not change the character of the network.

## 3. Description of networks analyzed

*Sources of weighted networks*

The weighted networks used in the present study were obtained from numerous sources. Neuronal avalanche networks were obtained from our own laboratory, DSI and fMRI networks were obtained from the group of Olaf Sporns, but the majority of the networks were obtain from the following three sources/databases: 1) the data provided by Mark EJ Newman at http://www-personal.umich.edu/~mejn/netdata/ , 2) Pajek (a software program for large network analysis) website, originally found at http://vlado.fmf.uni-lj.si/pub/networks/pajek/ but has since migrated to http://pajek.imfm.si/doku.php , 3) University of Florida Sparse Matrix Collection (UFSMC) at http://www.cise.ufl.edu/research/sparse/matrices/ which is also hosted at http://aws.amazon.com/datasets/Mathematics/2379. One can also search this database at http://www2.research.att.com/~yifanhu/GALLERY/GRAPHS/search.html or at http://aws.amazon.com/datasets/Mathematics/2379. We now provide more details about each of the networks (network groups) studied.

*Weighted neuronal avalanche networks* (Avalanche networks, n=3)

Functional cortical architectures of neuronal avalanches represent weighted directed networks derived as described previously [12]. In short, spontaneous synchronized activity was recorded in organotypic cortex slices cultured on integrated, planar 8x8 multi-electrode arrays (MEA) [13]. The local voltage fluctuations at each electrode site was thresholded and the time series of suprathreshold events at each electrode was taken as node activations in a 60-node networks (corner electrodes were not present). Cascades of node activations have been shown to form spatiotemporal clusters whose size distributions obey a power law with slope of -1.5, the hallmark of neuronal avalanches. By observing the spatio-temporal evolution of node activities on the network, a directed, weighted graph is derived [12]. The first data set was based on 7 cultures with stationary avalanche rate [13]. In a second data set, avalanche rate changed by an order of magnitude due to external slow driving [14]. We used three data sets to study functional neuronal avalanche connectivity in awake macaque monkeys. The first data set was derived from monkey 1 described in [15] based on ongoing avalanche activity in premotor cortex (N = 32 microelectrodes). The 2$^{nd}$ and 3$^{rd}$ data sets were obtained in 2 other awake, quietly sitting macaque monkeys (NIMH) by recording ongoing avalanche activity in the premotor cortex with 10x10-electrode arrays (0.6 mm interelectrode distance; N ~ 100). Functional architectures were reconstructed as described in [12] using a time step of 2 ms and an LFP threshold of -2.5 SD of signal fluctuations.

*Structural and functional human cortex core* (Human Brain, n=2)

The structural and functional connectivity data of the human cerebral cortex from the same 5 human subjects was recently published [16] and is available at http://www.indiana.edu/~cortex/resources.html. The nodes in these networks represent cortical regions of interests (N = 998) distributed over 67 functional cortical areas. The structural human cortex core has been identified using diffusion spectrum imaging (DSI) which includes ~15,000 fiber bundles of various densities that reflect the connection capacity between regions. The functional connectivity was based on correlations in the resting BOLD signal of fMRI between the same N=998 cortical regions of interest. Since such a network is fully connected (complete), we obtained the sparse functional networks by keeping only those links for which pair-wise correlations *R* in the fMRI signal were larger than 0.2.

*Gene regulatory networks* (Gene, n=3)

We used two human gene regulatory networks (N≅ 22300 and 14300) and one mouse network (N≅45100). They were obtained from the University of Florida Sparse Matrix Collection (UFSMC), posted by Vicenzo Belcastro's group, and described in [17]. Nodes in these networks represent individual genes and the links between them relate the expression level of each gene with the expression of other genes. The weights do not represent correlations, but rather a value of a parameter value in ODE-based algorithm, NIR [18]. Due to the large size of these networks and a very large number of significant links, we studied networks that were either sub-sampled versions of the original networks (see Suppl Figure 1 for networks sub-sampled at N=1000 and N=2000 nodes) or in which only links with an interaction parameter greater than 0.08 were kept, yielding very sparse networks that could be analyzed in a reasonable amount of time. Either method led to the same conclusion in terms of our pruning and link clustering analysis. The sub-sampled versions produced very similar results and were robust even if only 1000 or 2000 nodes were used in subsampling (see Suppl. Fig. S1). The results were also similar to those of the thresholded networks with the full set of nodes shown in Fig. 1F.

*Actor Collaboration Network* (Actor, n=1)

The actor networks were reconstructed using data from the Internet Movie Database (IMDb), provided by the Pajek Group provided in a Matlab format on the Pajek website [http://pajek.imfm.si/doku.php](http://pajek.imfm.si/doku.php). The original data contained a bipartite graph connecting 428K movies to 896K actors that were participating in them. From this bipartite graph, we reconstructed a weighted network in which the nodes represent actors and the link weights represent the number of movies in which they appeared together. To make this network computationally manageable, we first only considered movies with more than 5 actors in it and for the following categories: Drama, Short Documentary, Comedy, Western, Family, Mystery, Thriller, Music, Crime, Sci-Fi, Horror, War, Fantasy, Romance, Adventure, Animation, Action, Musical, Film-Noir. Second, we only kept actors who appeared in at least 10 movies. The final network had N=53K nodes and its properties are listed in Table I.

*"Les Miserables" Characters network* (Les Miserables, n=1)

The co-appearance network of characters in the novel *Les Miserables* has 77 nodes and weights represent the number of chapters in which a pair of characters appeared together. This network was originally created and studied in [19] and was obtained from the MEJ Newman web-site (http://www-personal.umich.edu/~mejn/netdata/).

*Words co-occurrence and Free Association Networks* (Words, n=2)

We used two different word networks. In the word co-occurrence *Reuters News 9/11* network, nodes represent keywords that occurred together in *Reuters News* articles on September 11, 2001, the day of the terrorist attacks in USA. The link weights represent the frequency of their co-occurrence. Originally produced by Steve Corman and Kevin Dooley at Arizona State University, the data are publicly available at http://pajek.imfm.si. The *Free Association Word network* (FA Word) is a directed network, in which source nodes represent normed words/cues to which >6,000 participants were asked to write the first word, the target node, that came to mind that was meaningfully related or strongly associated to the presented word,. The mechanics of this survey consists of a long list of words with the blank shown next to each item. For example, if given BOOK _________, they might write READ on the blank next to it. This procedure is called a discrete association task because each participant is asked to produce only a single associate to each word. This network can be found on the *Pajek* (http://vlado.fmf.uni-lj.si/pub/networks/data/dic/fa/FreeAssoc.htm), or USF website (http://w3.usf.edu/FreeAssociation/AppendixA/index.html).

For additional details see also http://w3.usf.edu/FreeAssociation/Intro.html.

*Caenorabditis elegans* (C.-elegans) Network

The neural system of the nematode worm *C. elegans* is comprised of a total of N = 302 neurons, most of which are linked together into one large, network. Our calculations are based on three versions of this network. We used a recently improved *C. elegans* neuronal data base [20] that contains one network based on chemical, i.e. synaptic, connections and one network based on electrical, i.e. gap-junction, mediated connections between neurons (available at http://mit.edu/lrv/www/elegans/). Link weights in these networks represent multiplicity of connections between neurons. For comparison, we also analysed an earlier version of this network [21] with its small-worldness introduced in [1] and which is available at http://www-personal.umich.edu/~mejn/netdata/. Results for all three networks did not differ substantially and were averaged for presentation purposes.

*Scientific author collaboration networks* (Collaboration Networks, n=4)

In author collaboration networks, authors from different disciplines in physics represent nodes and are connected, if they co-author a paper. Link weights in these networks quantify the number of papers co-authored, each paper carrying the weight inversely proportional to the total number of the authors. The disciplines 'Condensed Matter', 'Network Sciences', 'High Energy Physics', and 'Astrophysics' with N = 1,500 – 17,000 authors, i.e. nodes, were analysed (available at http://www-personal.umich.edu/~mejn/netdata/).

*Airline transportation network* (Transportation Networks, n=2)

The US Air airline network is an undirected, weighted transportation network with N = 332 nodes representing airports around the world. Link weights represent the relative number of the flights US Air had in 1997 (http://www.cise.ufl.edu/research/sparse/matrices/Pajek/USAir97.html). We also used an airport network (http://wiki.gephi.org/index.php/Datasets) in which the nodes constitute 500 airports in the US and link weights represent the number of passengers transported each year.

*Weighted Evolving Networks*

We created networks based on two popular models of weighted evolving networks, i.e., in which weights are assigned during growth as nodes and links are added. The two growth models (GM) assign weights according to (1) resources reserved based on the degree of the connecting node [22] or (2) fixed resources distributed based on the relative node strengths [23]. These two rules were originally applied to preferential attachment models and as such did not produce networks with any excess clustering. We therefore applied the corresponding weight assignment rules to the OHO [11] growing network, which has significant $\Delta C$ and named them (1) GM1 OHO and (2) GM2 OHO. The third growth model was based on the simulations by Kumpula et al. [24], which uses local neighbourhood searches to increase the number common neighbours and corresponding link weights. The critical model parameter is the relative weight increase for closed triangles delta, which we studied for 5 different values within the of zero to 1.

## 4. Supplementary Table

Table 1: *Summary of network properties.* The first column contains the network name and the number of actual networks analyzed in parenthesis. The data columns are as follows: *N*: number of network nodes. <*k*>: mean node degree. <*d*>: mean network diameter. $r_A$ : assortativity based on degree-degree correlations. *C*: average node clustering coefficient. *ΔC*: mean excess clustering. *Q*: Network modularity obtained using Girvan-Newman algorithm [25,26]. *M* and $R_{C_L}$ as defined in the main text.

| Networks (#nets) | N | $\langle k \rangle$ | $\langle d \rangle$ | $r_A$ | C | $\Delta C$ | Q | M | $R_{C_L}$ |
|---|---|---|---|---|---|---|---|---|---|
| **Neural** | | | | | | | | | |
| DSI Human Brain (5) | 998 | 36 | 3.1 | 0.29 | 0.47 | 0.42 | 0.68 ± 0.08 | 0.34 | 0.59 |
| fMRI Human Brain (5) | 998 | 67 | 2.7 | 0.25 | 0.53 | 0.44 | 0.62 | 0.56 | 0.63 |
| Avalanche Monkey (3) | 77 ± 14 | 13 ± 4 | 2.6 ± 0.19 | 0.3 ± 0.13 | 0.51 ± 0.03 | 0.29 ± 0.03 | 0.48 ± 0.03 | 0.39 | 0.45 |
| Avalanche Culture (7) | 59 | 16 ± 3 | 2.3 ± 0.17 | 0.28 ± 0.11 | 0.63 ± 0.03 | 0.32 ± 0.03 | 0.5 ± 0.06 | 0.65 | 0.31 |
| Aval. Culture Driven (7) | 58 ± 1 | 16 ± 4 | 2.2 ± 0.16 | 0.24 ± 0.11 | 0.57 ± 0.03 | 0.26 ± 0.03 | 0.41 ± 0.07 | 0.72 | 0.22 |
| C-elegans (3) | 285 ± 10 | 7.9 | 3.6 ± 0.3 | 0.02 ± 0.07 | 0.23 ± 0.01 | 0.15 ± 0.01 | 0.5 ± 0.04 | 0.4 | 0.002 |
| **Transportation** | | | | | | | | | |
| US Air (1) | 332 | 13 | 2.7 | -0.21 | 0.75 | 0.24 | 0.2 | -0.19 | -0.025 |
| US airports (1) | 500 | 12 | 3 | -0.27 | 0.73 | 0.18 | 0.28 | -0.059 | 0.31 |
| **Human** | | | | | | | | | |
| Actors (1) | 53960 | 6.6 | 7.6 | 0.18 | 0.58 | 0.53 | 0.68 | 0.47 | 0.36 |
| Les Miserables (1) | 77 | 6.6 | 2.6 | -0.16 | 0.74 | 0.47 | 0.53 | 0.25 | 0.16 |
| **Genes** | | | | | | | | | |
| Human Gene 1 (1) | 22282 | 15 | 5.3 | 0.068 | 0.66 | 0.52 | 0.69 | 0.59 | 0.53 |
| Human Gene 2 (1) | 14337 | 19 | 3.6 | -0.0047 | 0.65 | 0.46 | 0.6 | 0.56 | 0.55 |
| Mouse Gene (1) | 45101 | 5.5 | 4.9 | 0.3 | 0.57 | 0.51 | 0.74 | 0.6 | 0.59 |
| **Language** | | | | | | | | | |
| Reuters News 9/11 (1) | 13314 | 22 | 3.1 | -0.11 | 0.39 | 0.22 | 0.24 | 0.2 | 0.14 |
| Language Free Assoc. (1) | 10617 | 6.8 | 4.8 | -0.0076 | 0.13 | 0.12 | 0.52 | 0.37 | 0.23 |
| **Collaboration** | | | | | | | | | |
| Condensed Matter (1) | 16726 | 5.7 | 6.6 | 0.18 | 0.74 | 0.74 | 0.52 | -0.16 | -0.19 |
| High Energy Physics (1) | 8361 | 3.8 | 7 | 0.29 | 0.64 | 0.63 | 0.52 | -0.091 | -0.13 |
| Astrophysics (1) | 16706 | 15 | 4.8 | 0.24 | 0.73 | 0.72 | 0.53 | -0.04 | -0.19 |
| Network Science (1) | 1589 | 3.5 | 5.8 | 0.46 | 0.88 | 0.87 | 0.61 | -0.042 | -0.43 |
| **Learning** | | | | | | | | | |
| LSCrit (10) | 60 | 11 | 2.3 ± 0.03 | 0.12 ± 0.04 | 0.68 | 0.44 | 0.78 ± 0.06 | 0.15 | 0.27 |
| LSSub (10) | 60 | 11 | 2.3 ± 0.04 | 0.15 ± 0.02 | 0.67 | 0.43 | 0.79 ± 0.02 | 0.17 | 0.016 |
| LSSup (10) | 60 | 11 | 2.3 ± 0.02 | 0.15 ± 0.03 | 0.68 | 0.44 | 0.64 ± 0.14 | 0.075 | 0.25 |
| ASCrit (10) | 60 | 11 | 2.3 ± 0.05 | 0.15 ± 0.03 | 0.68 | 0.44 | 0.44 ± 0.04 | -0.1 | -0.059 |
| ASSub (10) | 60 | 11 | 2.3 ± 0.03 | 0.15 ± 0.03 | 0.68 | 0.44 | 0.33 ± 0.02 | -0.11 | -0.21 |
| ASSup(10) | 60 | 11 | 2.3 ± 0.04 | 0.15 ± 0.04 | 0.68 | 0.44 | 0.47 ± 0.01 | -0.0011 | 6.3e-16 |
| **Models** | | | | | | | | | |
| OHO Type I (10) | 100 | 12 | 2.7 ± 0.06 | 0.2 ± 0.015 | 0.67 | 0.52 | 0.57 ± 0.01 | -0.0023 | -0.011 |
| OHO Type II (10) | 100 | 12 | 2.7 ± 0.07 | 0.2 ± 0.016 | 0.67 | 0.52 | 0.5 ± 0.01 | -0.41 | -0.41 |
| OHO Type III (10) | 100 | 12 | 2.7 ± 0.07 | 0.19 ± 0.03 | 0.67 | 0.52 | 0.65 ± 0.01 | 0.58 | 0.55 |
| WN Type I (10) | 100 | 9.8 | 2.5 | -0.03 ± 0.03 | 0.46 | 0.37 | 0.56 ± 0.01 | 0.0063 | -0.005 |
| WN Type II (10) | 100 | 9.8 | 2.5 | -0.01 ± 0.04 | 0.46 | 0.37 | 0.54± 0.01 | -0.4 | -0.13 |
| WN Type III (10) | 100 | 9.8 | 2.5 ± 0.05 | 0.05 ± 0.04 | 0.46 | 0.4 | 0.6 ± 0.01 | 0.4 | 0.48 |
| **Growth Models** | | | | | | | | | |
| GM1 OHO $W_1$ (10) | 100 | 12 | 2.7 ± 0.03 | 0.19 ± 0.02 | 0.67 | 0.52 | 0.5 ± 0.01 | -0.46 | -0.3 |
| GM2 OHO $W_2$ (10) | 60 | 11 | 2.3 ± 0.04 | 0.2 ± 0.03 | 0.67 | 0.43 | 0.5 ± 0.01 | 0.038 | -0.17 |

## 5. Supplementary Figures

**Figure S1**. Definition of link clustering and excess clustering. **a**, Link clustering coefficient $C_L$ defined as the relative overlap between neighbourhoods of the link's end nodes. $n_C$: common nodes; $n_U$: uncommon nodes. Note that neighbourhood in directed networks can be defined based on incoming ('in'), outgoing ('out'), or all ('all') links for each node. If not stated otherwise, we use 'out' neighbourhoods for all analysis. For clarity, node indices have been used. **b**, Excess clustering. *Left*: Bottom-pruning analysis of the clustering coefficient $C$ for n = 7 weighted, directed functional neuronal avalanche networks. *Right*: Single example of a bottom-pruned network at f=0.3 and 0.9 indicated by red arrows in left panel. *DSPR*: degree-sequence preserved randomization. $\Delta C$: excess node clustering.

**Figure S2**. Integrative weight organization for gene networks is also obtained when reducing network size by random node sub-sampling instead of removal of weakest links, e.g. 0.08 threshold used in figure 1. **a**, Results obtained by sub-sampling N = 2,000 nodes from the original gene networks. For each genome, five sub-sampled networks were averaged and their link clustering analysis (*left*) and pruning analysis (*right*) are shown. **b**, Corresponding analysis for sub-sampling N = 1,000 nodes (10 subsamples averaged for each genome).

**Figure S3**. Scatter plot of $M$ and $R_{C_L}$ for directed weighted networks (see Suppl. Table and Fig. 2i main text for details) using different definitions of neighbourhood. *Src*:

Source node; *Targ*: Target node. We see that $R_{C_L}$ is very robust to the choice of neighbourhood, while *M* shows greater variability. Nevertheless, classification of networks into dispersive, neutral, and integrative is fairly robust to neighbourhood definition. For network legend see Fig. 2i main text and Fig. S8.

**Figure S4**. Changes in *ΔC* are the result of changes in $C_{orig}$, and $C_{DSPR}$, so the same change in excess clustering can be obtained in many different ways. To obtain a more detailed picture, the same networks and weight organizations as in Figure 3 in the main text are shown here with both $C_{orig}$ (solid lines) and $C_{DSPR}$ (broken lines) plotted separately, with left panels showing bottom-pruning and right top-pruning.  **a**, Bottom pruning (left) and top pruning (right) of OHO (black) and WN (blue) neutral networks (solid lines) and corresponding DSPR controls (broken lines; n = 10).  Note linear decay as predicted by theory for both the original and randomized controls.  **b**, For bottom pruning, *C* remains relatively high in this dispersive network model, but the increase in $C_{DSPR}$ leads to an overall reduction in *ΔC*, particularly for OHO topology.  **c**, Conversely, $C_{DSPR}$ increases for top- but not bottom pruning in integrative networks.  The symmetry between the integrative vs. dispersive and bottom vs. top pruning for OHO topology is the result of its inverse linear relationship between the node clustering coefficients and degrees. In most topologies, node clustering coefficient and node degree are inversely related, $C \sim k^{-\beta}$, with $0<\beta<1$ and thus we expect similar results in other topologies.

**Figure S5**. Same simulations as in figure 3 main text, but now using the "all" definition for the neighbourhood in directed networks, showing virtually the same results. **a**,

Neutral networks with independent link weights implemented on OHO (*black*) and WN (*red*) topologies. *Left*: $\Delta C_L$ does not correlate with weight rank. *Right*: $\Delta C$ decreases monotonically with $f$ for bottom- (*solid lines*) and top-pruning (*broken lines*). **b**, Corresponding analysis for dispersive networks where $w_{ij}$ are assigned as the geometric mean of the end-node degrees $k_i$ and $k_j$. **c**, Corresponding analysis for integrative networks. See main text Figure 3 for further details. Simulations of n=10 networks each (N=100; <k> = 12,10 for OHO, WN).

**Figure S6**. Link clustering and robustness to pruning for growth model 1 (**a**) and growth model 2 (**b**).

**Figure S7**. We applied the weighting from Eq. S5 that aims to de-correlate our two measures of weight organization, $M$ and $R_{C_L}$. **a**, Results for OHO topology using 16 different values for parameter *a* in Eq. S5, approximately logarithmically spaced between 0.01 and 100. Three different network sizes were used N=60 (black), N=100 (blue), N=200 (red). **b,** the same results as in *a*, applied to WN topology.

**Figure S8**. Comparison of $R_{C_L}$ with alternative network measures. **a**, $R_{C_L}$ correlates with $R_w$ as described in Eq. S4. Thus, link weights organize along two different local measures of clustering, the relative fraction of common neighbours, and the clustering coefficients of the corresponding end nodes. **b**, The local measure $R_{C_L}$ weakly correlates

with modularity Q, a global measure of community structure, which in the current analysis takes link weight into account (R=0.35) [25].

Reference List

1. Watts,D.J. & Strogatz,S.H. Collective dynamics of 'small-world' networks. *Nature* **393**, 440-442 (1998).

2. Radicchi,F., Castellano,C., Cecconi,F., Loreto,V. & Parisi,D. Defining and identifying communities in networks. *Proc. Natl. Acad. Sci. U. S. A* **101**, 2658-2663 (2004).

3. Maslov,S. & Sneppen,K. Specificity and stability in topology of protein networks. *Science* **296**, 910-913 (2002).

4. Newman,M.E.J. Properties of highly clustered networks. *2003* **68**, 026121 (2003).

5. Ravasz,E. & Barabasi,A.L. Hierarchical organization in complex networks. *Phys Rev. E. Stat. Nonlin. Soft. Matter Phys* **67**, 026112 (2003).

6. Dorogovtsev,S.N. & Mendes,J.F.F. Evolution of networks: From biological nets to the internet and WWW. University Press, Oxford, USA (2003).

7. Dorogovtsev,S.N. & Mendes,J.F.F. Evolution of networks. *Advances in Physics* 1079-1187 (2002).

8. Boguna,M., Krioukov,D. & Claffy,K.C. Navigability of complex networks. *Nature Physics* **5**, 74-80 (2008).

9. Serrano,M.A., Krioukov,D. & Boguna,M. Self-similarity of complex networks and hidden metric spaces. *Phys Rev. Lett.* **100**, 078701 (2008).

10. Newman,M.E. & Watts,D.J. Scaling and percolation in the small-world network model. *Phys Rev. E. Stat. Phys Plasmas. Fluids Relat Interdiscip. Topics.* **60**, 7332-7342 (1999).

11. Ozik,J., Hunt,B.R. & Ott,E. Growing networks with geographical attachment preference: emergence of small worlds. *Phys Rev. E. Stat. Nonlin. Soft. Matter Phys* **69**, 026108 (2004).

12. Pajevic,S. & Plenz,D. Efficient network reconstruction from dynamical cascades identifies small-world topology from neuronal avalanches. *PLoS Comp. Biol.* **5**, e1000271 (2008).

13. Beggs,J.M. & Plenz,D. Neuronal avalanches in neocortical circuits. *J. Neurosci.* **23**, 11167-11177 (2003).

14. Plenz,D. & Chialvo,D.R. Scaling properties of neuronal avalanches are consistent with critical dynamics. *arXiv:0912. 5369* (2010).

15. Petermann,T. *et al.* Spontaneous cortical activity in awake monkeys composed of neuronal avalanches. *Proc. Natl. Acad. Sci. U. S. A* **106**, 15921-15926 (2009).

16. Hagmann,P. *et al.* Mapping the structural core of human cerebral cortex. *PLoS. Biol.* **6**, e159 (2008).

17. Gregoretti,F., Belcastro,V., di Bernardo,D. & Oliva,G. A parallel implementation of the network identification by multiple regression (NIR) algorithm to reverse-engineer regulatory gene networks. *PLoS ONE.* **5**, e10179 (2010).

18. Gardner,T.S., di Bernardo,D., Lorenz,D. & Collins,J.J. Inferring genetic networks and identifying compound mode of action via expression profiling. *Science* **301**, 102-105 (2003).

19. Knuth,D.E. The Stanford GraphBase: A Platform for Combinatorical Computing. Addison-Wesley, Reading, MA (1993).

20. Varshney,L.R., Chen,B.L., Paniagua,E., Hall,D.H. & Chklovskii,D.B. Structural properties of the *Caenorhabditis elegans* neuronal network. *arXiv:0907. 2373v2* (2010).

21. White,J.G., Southgate,E., Thomson,J.N. & Brenner,S. The structure of the nervous system of the nematode Caenorhabditis elegans. *Phil. Trans. R. Soc. Lond.* **314**, 1-340 (1986).

22. Yook,S.H., Jeong,H., Barabasi,A.L. & Tu,Y. Weighted evolving networks. *Phys. Rev. Lett.* **86**, 5835-5838 (2001).

23. Barrat,A., Barthelemy,M. & Vespignani,A. Weighted evolving networks: coupling topology and weight dynamics. *Phys Rev. Lett* **92**, 228701 (2004).

24. Kumpula,J.M., Onnela,J.P., Saramaki,J., Kaski,K. & Kertesz,J. Emergence of communities in weighted networks. *Phys Rev. Lett.* **99**, 228701 (2007).

25. Girvan,M. & Newman,M.E. Community structure in social and biological networks. *Proc. Natl. Acad. Sci. U. S. A* **99**, 7821-7826 (2002).

26. Newman,M.E.J. Analysis of weighted networks. *Phys. Rev. E.* **70**, 056131 (2004).

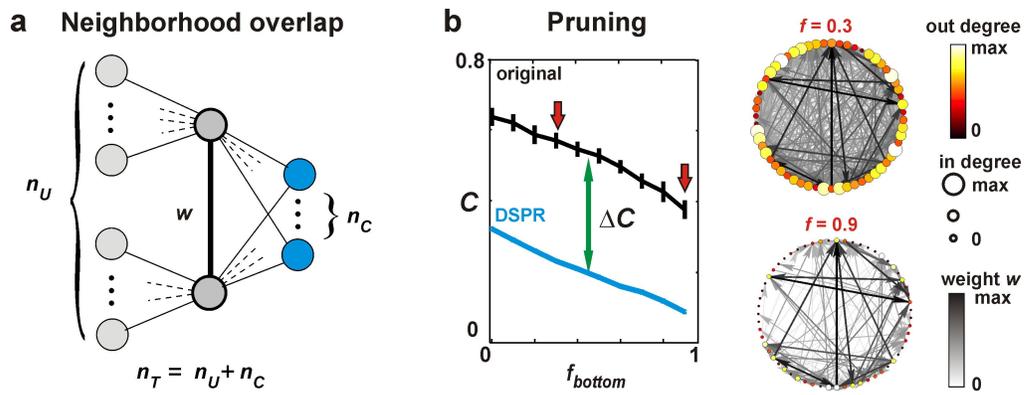

*Pajevic & Plenz*
*Figure S1*
*Methods*
*2 cols*

**a  Gene interactions**
— human genome 1
— human genome 2
— mouse genome

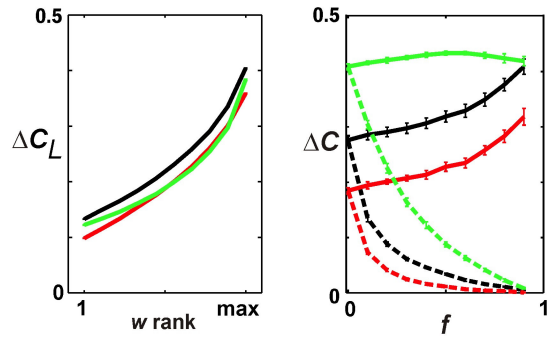

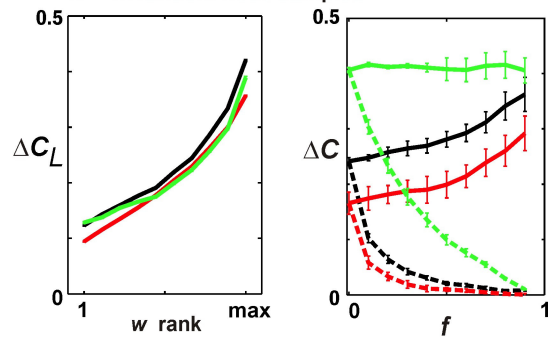

*Pajevic &Plenz*
*Figure S2*
*Gene Subsample*
*1 col*

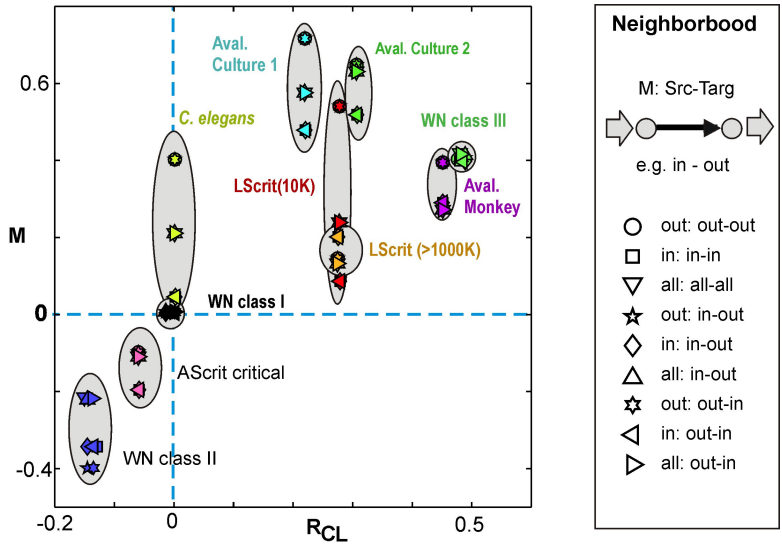

*Pajevic &Plenz*
*Figure S3*
*NN dependency*
*1.5 col*

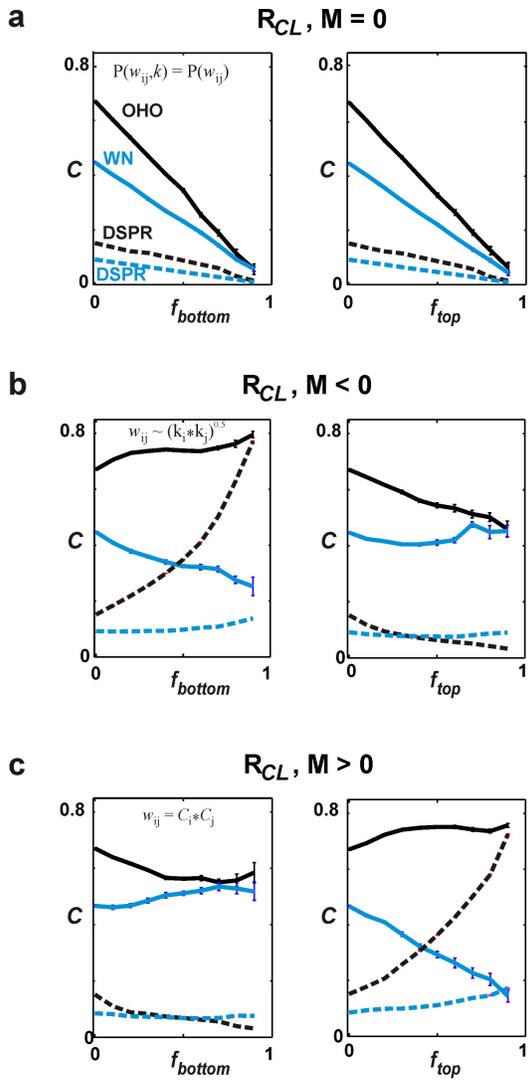

*Pajevic & Plenz*
*Figure S4*
*Model C*
*2 col*

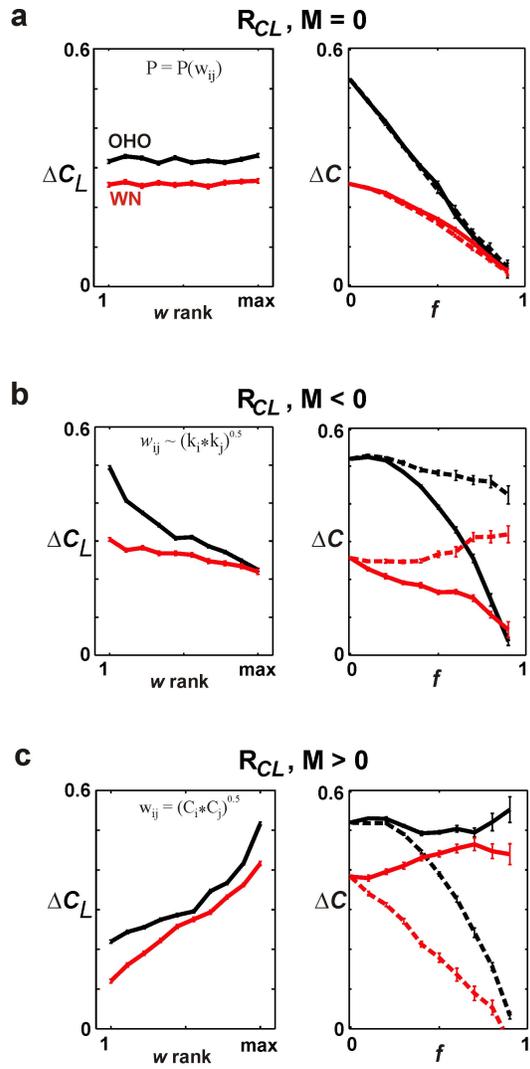

*Pajevic & Plenz*
*Figure S5*
*All NN*
*2 col*

**a**   GM1

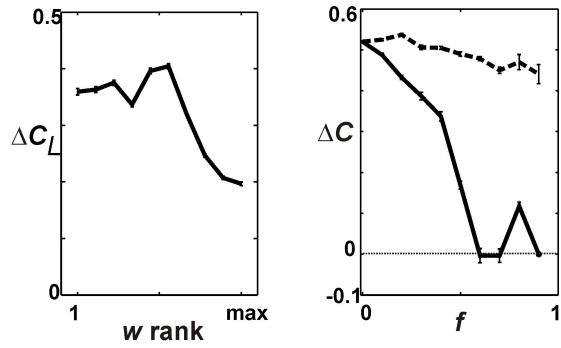

**b**   GM2

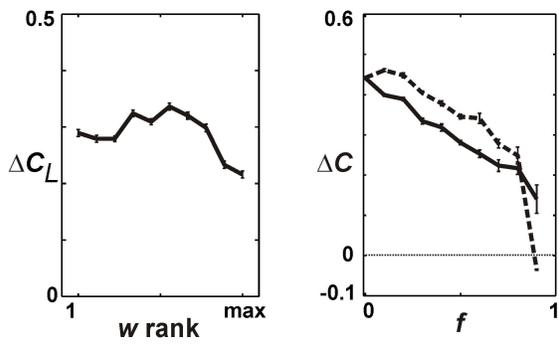

*Pajevic &Plenz*
*Figure S6*
*Growth models*
*1 col*

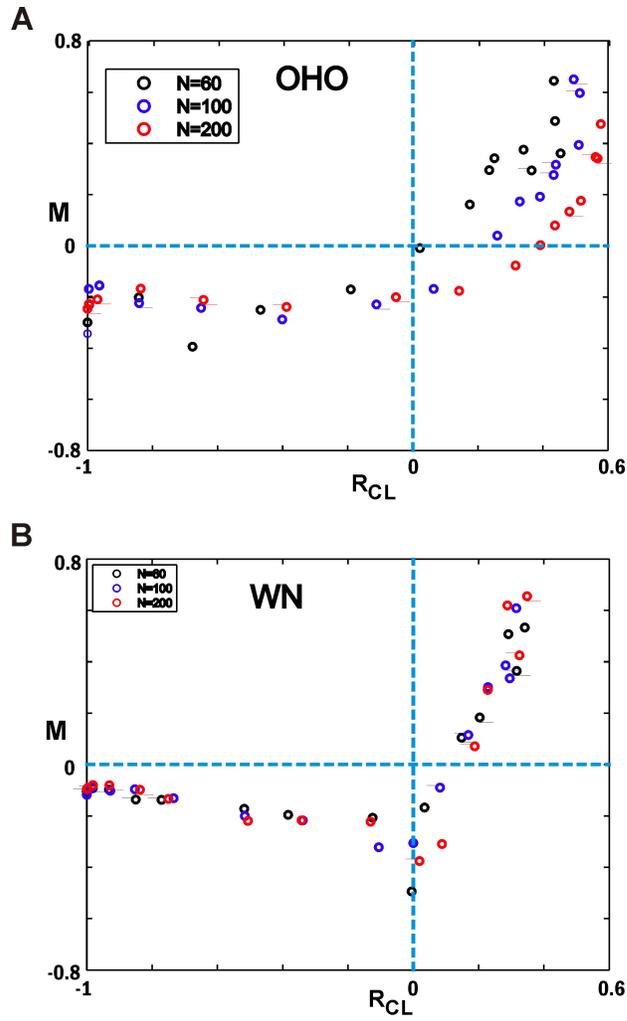

*Pajevic &Plenz*
*Figure S7*
*Decorr Rcl & M*
*1 col*

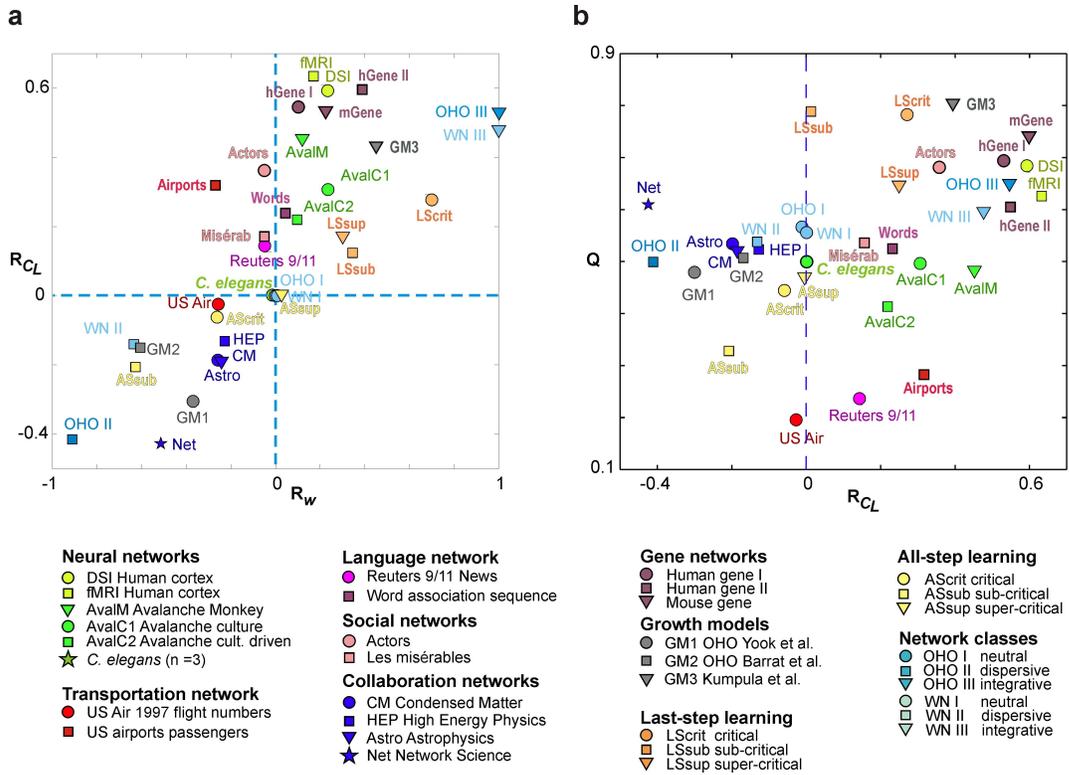

*Pajevic & Plenz*
*Figure S8*
*Rw and Modulatrity*
*2 cols*